\documentclass[aps,reprint,superscriptaddress,longbibliography]{revtex4-1}
\usepackage[colorlinks,bookmarks=false,citecolor=red,linkcolor=blue,urlcolor=blue]{hyperref}
\usepackage{amsmath,amssymb,graphicx,color}
\usepackage{booktabs}
\usepackage{bm} 
\usepackage{amssymb}
\usepackage{color}
\usepackage{multirow}
\usepackage{mathrsfs,bbm}
\usepackage{amsmath}
\usepackage{verbatim}
\usepackage{bbm}
\usepackage[version=3]{mhchem}
\usepackage{array}
\newcolumntype{P}[1]{>{\centering\arraybackslash}p{#1}}
\usepackage{epstopdf}
\usepackage{wrapfig}

\def\veps{\varepsilon}
\def \r{{\bf r}}

\definecolor{mycol}{rgb}{0,0,0}  
\begin{document}
\title{Andreev and normal reflections in gapped bilayer graphene-superconductor junctions}
\author{Panch Ram}
\email{panch.ram@uni-konstanz.de}
\affiliation{Fachbereich Physik, Universität Konstanz, D-78457 Konstanz, Germany}
\author{Detlef Beckmann}
\affiliation{Institute for Quantum Materials and Technologies, Karlsruhe Institute of Technology, Karlsruhe D-76021, Germany}
\author{Romain Danneau}
\affiliation{Institute for Quantum Materials and Technologies, Karlsruhe Institute of Technology, Karlsruhe D-76021, Germany}
\author{Wolfgang Belzig}
\email{wolfgang.belzig@uni-konstanz.de}
\affiliation{Fachbereich Physik, Universität Konstanz, D-78457 Konstanz, Germany}

\date{\today}	
\begin{abstract}
We study the Andreev and normal reflection processes---retro as well as specular---in a bilayer graphene-superconductor junction where equal and opposite displacement fields are applied for the top and bottom layers to induce a band gap. By employing the Dirac-Bogoliubov-de Gennes equation for the gapped bilayer graphene-superconductor junction, we calculate the reflections probabilities within the scattering theory approach. The subgap conductance, calculated in the framework of Blonder-Tinkham-Klapwijk formalism, shows the contribution from the Andreev retro-reflection (specular-reflection) when the applied bias voltage is below (above) the Fermi energy. Notably, both retro and specular reflections are modified in the presence of the displacement field, and the retro-to-specular crossover gets amplified when the displacement field is relatively small. They can be further tuned to either specular or retro Andreev reflection by adjusting the Fermi energy. Furthermore, our study reveals the simultaneous existence of double Andreev reflections and double normal reflections when the displacement field becomes comparable to the interlayer coupling strength. The existence of the normal retro-reflection process in a bilayer graphene-superconductor junction is a new finding which shows a distinctive feature in the conductance that can be experimentally verified.
\end{abstract}
\maketitle
\section{Introduction}
Andreev reflection (AR) is a scattering process that occurs at a normal-superconductor (NS) junction and is solely responsible for converting a dissipative normal current into a dissipationless supercurrent~\cite{Andreev1964,Tinkham2004}. In this process, when an electron from the N side is incident at the junction with excitation energy ($\veps$) less than the superconducting gap ($\Delta$), it is reflected back as a hole in a retro-reflection manner and a charge $2e$ is transferred on S side as a Cooper pair~\cite{BCS1957}. The effect of retro Andreev reflection (RAR) on the current-voltage relation has been studied in the seminal paper of the so-called BTK theory~\cite{Btk1982, Shelankov1982}. One consequence of it is the existence of subgap conductance for bias voltage $eV<\Delta$ which can attain twice the value of the normal state conductance for a perfect transparent junction. Experimentally, a direct effect of RAR has also been observed as a sign change of velocity upon reflection~\cite{Benistant1983} and zero-bias anomaly~\cite{Kastalsky1991}.

Graphene is an interesting system which has attracted the attention of the condensed matter community, not only due to its distinctive transport properties~\cite{Novos2005, Zhang2005, Novos2006, Katsnelson2006}, owing to the relativistic linear dispersion and vanishing density of states at the Dirac points, but also because of a new phenomenon of electron-hole conversion at the graphene-superconductor junction---specular Andreev reflection (SAR)~\cite{Bee2006,Beenakker2008}. The nature of AR at the junction is expected to change from retro to specular since the Fermi energy ($E_F$) in graphene is significantly lower than the conventional metals. More precisely, when the Fermi energy is close to the charge neutrality point (CNP), i.e. $E_F\to0$, the incident electron from the conduction band (above the CNP) is reflected (in a specular manner) as a hole from the valence band (below the CNP). This interband electron-hole conversion phenomenon is known as the SAR. The study of SAR~\cite{Bee2006,Beenakker2008} has prompted a surge of interest in the graphene-superconductor junction, and thereafter various theoretical works have been put forward~\cite{Bhatta2006,Linder2007,Linder2008,Zhang2008,Benjamin2008,Majidi2012}. But, the experimental studies of Andreev processes have been  limited so far \cite{Popinciuc2012,Han2018,Pandey2019,Bhandari2020,Pandey2021,Pandey2022,Jois2023} and the observation of SAR in graphene has not been succeeded yet~\cite{Sahu2016} as the Fermi-energy fluctuation is $\delta E_F>\Delta$ which prevents attaining the $E_F\to0$ limit. 

In contrast, the bilayer graphene (BLG) is regarded as a more suitable system compared to the monolayer graphene for observing the SAR, due to its low $\delta E_F$ fluctuation~\cite{Efetov2016,Efetov_PRB2016,Ludwig2007,Takane2017}. Indeed, an experiment has been performed~\cite{Efetov2016} and the measured subgap differential conductance shows a small dip around the bias $eV=E_F$ when $E_F$ is tuned below $\Delta$. This characteristic feature has been accredited to a crossover from retro-to-specular Andreev reflection. However, the retro-to-specular crossover color plots show a very weak consent between the experiment and theory that are presented in Figs.~2 and~3(a) in Ref.~\cite{Efetov2016}, see especially the different colorbar scales used for the experimental and theoretical data plots. This raises speculation regarding the observation of SAR in BLG. Moreover, a theoretical study to enhance the SAR contribution in subgap conductance has been suggested by introducing a Zeeman field on the normal side of the bilayer graphene NS junction~\cite{Soori2018}.

In this paper, we propose an alternate study to enhance the retro-to-specular crossover in BLG by applying two different displacement fields. Without loss of generality, we consider that the generated electrostatic potentials, due to the displacement fields, are $\lambda$ for the top layer and $-\lambda$ for the bottom layer which creates a gap $\sim 2\lambda$ in the BLG band structure when $\lambda\ll t_\perp$, where $t_\perp$ is being the interlayer coupling. Consequently, the subgap differential conductance across the NS junction vanishes and widens the retro-to-specular crossover, from a point at bias $eV=E_F$ to a whole range of bias $|E_F-\lambda|<eV<|E_F+\lambda|$ {\color{mycol}with width $2\lambda$}. {\color{mycol}This enhancement due to the displacement field would possibly help to observe the crossover experimentally as the boundary becomes wider}. In addition, we also investigate the reflection processes and the corresponding differential conductance when the displacement field becomes large and comparable to the interlayer coupling, i.e., $\lambda\lesssim t_\perp$. Interestingly, in this regime, four reflection processes exist simultaneously: specular normal reflection (SNR), retro normal reflection (RNR), specular Andreev reflection (SAR), and retro normal reflection (RAR). {\color{mycol} This occurs due to the partial inversion of the lower conduction/valence band upon the applied displacement field. The existence of RNR is a new finding in the system of bilayer graphene-superconductor junction which exhibits a distinctive characteristic feature in the differential conductance.} 

The rest of the paper is organized as follows. In Sec.~\ref{sec:BLG}, we present the effective low-energy Hamiltonian of the gapped bilayer graphene and the corresponding band structure. The model and formalism are illustrated in Sec.~\ref{sec:model-formalism}, introducing the Dirac-Bogoliubov-de Gennes (DBdG) equation and the excitation energy eigenstates for the normal and superconducting sides of the junction. Sec.~\ref{sec:result} provides the results for the reflection probabilities and normalized conductance. They are discussed in detail separately for the small and large displacement fields in subsections~\ref{subsec:small-lam},~\ref{subsec:large-lam}, and~\ref{subsec:exp}, respectively. Sec.~\ref{sec:conclusion} concludes the paper.
\section{Gapped bilayer graphene}\label{sec:BLG}
\begin{figure}[htb]
     \includegraphics[width=0.47\textwidth]{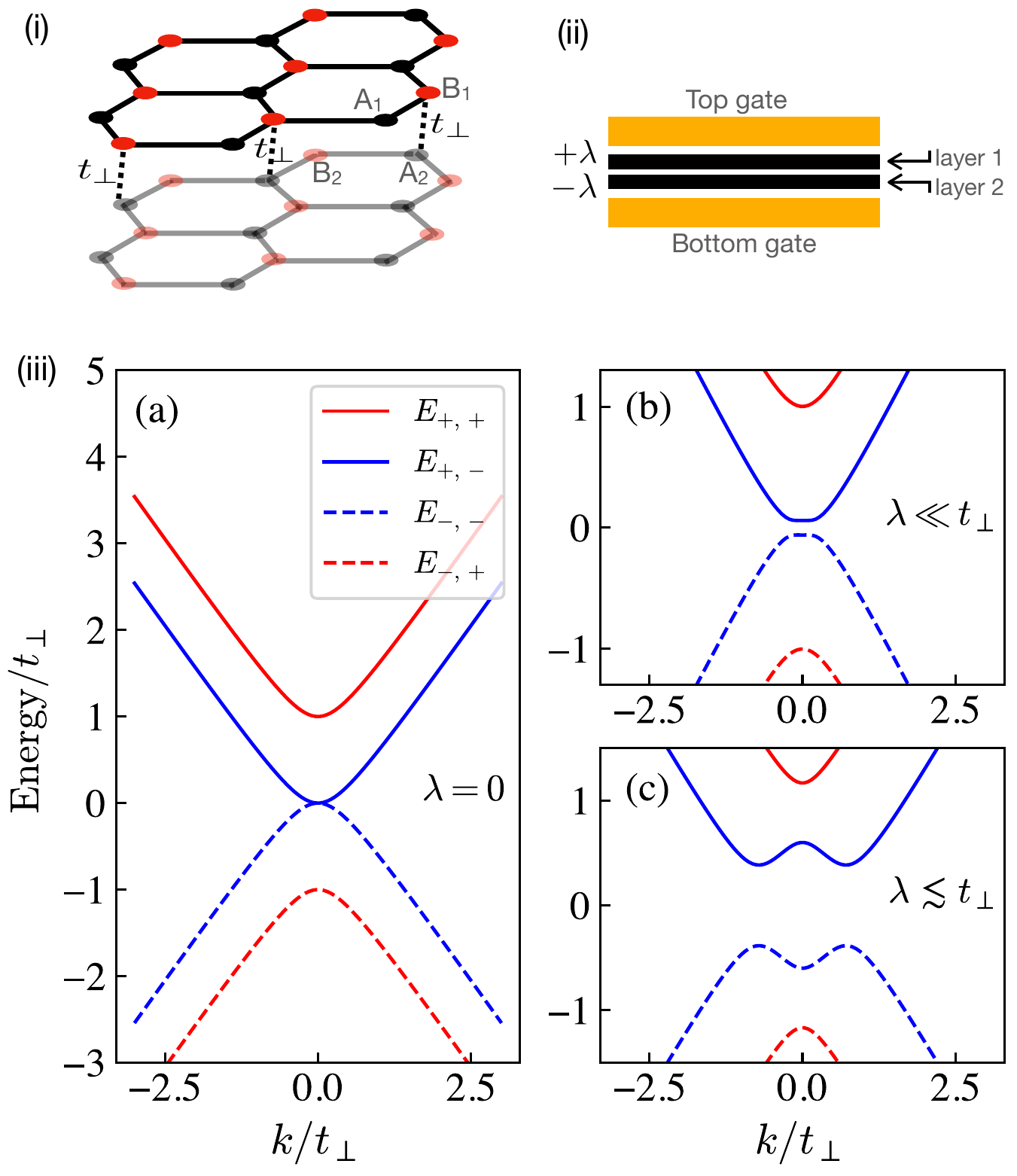}
    \caption{{\color{mycol}Bernal-stacked bilayer graphene structure and a schematic depiction to induce the displacement field are in (i) and (ii), respectively. In (iii),} the low-energy band structure of bilayer graphene, $E_{\nu,\pm}$ for $\nu=\pm$ in Eq.~\eqref{eq:bands}, with respect to $k$. The behavior of conduction bands ($E_{+,\pm}$) and valence bands ($E_{-,\pm}$) for the displacement field (a) $\lambda=0$, (b) $\lambda=0.06t_\perp$, and (c) $\lambda=0.6t_\perp$. We observe a band gap opening and the Mexican-hat-like shape formation in $E_{+,-}$ and $E_{-,-}$ as $\lambda$ is increased.}
    \label{fig:bands}
\end{figure}

The bilayer graphene comprises two layers of graphene, wherein two non-equivalent $A_1$ and $B_1$ carbon atoms in one layer and $A_2$ and $B_2$ in the other layer. The two layers are stacked either in $A_1$-$A_2$ or in {\color{mycol}$B_1$-$A_2$} structure~\cite{Yan2011,Brown2012}. A quantum Monte Carlo simulation suggests that the latter structure is more stable~\cite{Mostaani2015}, so we consider the {\color{mycol}$B_1$-$A_2$} stacking which is commonly known as the Bernal stacking~\cite{Yan2011}, {\color{mycol}see in Fig.~\ref{fig:bands}(i).} Within the tight-binding approximation and considering only the first nearest-neighbour hopping in each layer and {\color{mycol}$B_1$-$A_2$} interlayer hopping, the relevant low-energy Hamiltonian can be deduced in momentum-space~\cite{McCann2006,McCann2006prb,Barbier2009,Castro2009,McCann2013,Kraft2018v1,Wu2020} which reads in basis {\color{mycol}$(\psi_{A_1}~\psi_{B_1}~\psi_{A_2}~\psi_{B_2})^T$} as
{\color{mycol}\begin{align}
    H = \begin{pmatrix}
        \lambda &\hbar v_F k_{-} &0 &0 \\[0.2em]
        \hbar v_F k_{+} &\lambda &-t_\perp &0 \\[0.2em]
        0 &-t_\perp &-\lambda &\hbar v_F k_{-} \\[0.2em]
        0 &0 &\hbar v_F k_+ &-\lambda
    \end{pmatrix}
    \label{eq:H0}
\end{align}}
Here, {\color{mycol}$k_\pm = k_x\pm i\eta k_y \equiv -i(\partial_x \pm i\eta\partial_y)$} are the wave-vectors that are measured from the corners of $2$D  hexagonal Brillouin zone's Dirac points $K(K')$ for different valley $\eta=+(-)$; $v_F \simeq 10^6$ m/s is the Fermi-velocity and $\hbar$ is the reduced Planck's constant. $\lambda$ is added to account for the on-site potential (equal in magnitude but opposite in sign for the two layers) which can be tuned by different displacement fields. A schematic depiction to induce $\lambda$ is shown in {\color{mycol}Fig.~\ref{fig:bands}(ii).} 
The non-zero $\lambda$ creates a band gap in BLG band structure, and the gap can be tuned up to $250$ meV~\cite{Ohta2006,Oostinga2008,Zhang2009,Taychatanapat2010,Varlet2014,Kraft2018v2,Du2018}. We set $\hbar v_F= 1$ for the calculation hereafter.

The eigenvalues of $H$ in Eq.~\eqref{eq:H0} provide valley degenerated low-energy bands
\begin{align}
    E_{\nu,\pm} = \nu \sqrt{k^2 + \frac{t^2_\perp}{2} + \lambda^2  \pm \sqrt{\Big(\frac{t^2_\perp}{2}\Big)^2 + k^2\left(4\lambda^2+t^2_\perp\right) }}
    \label{eq:bands}
\end{align}
where $k=\sqrt{k_x^2+k_y^2}$ and the index $\nu=+(-)$ labels the conduction (valence) bands. In Fig.~\ref{fig:bands}(iii), we plot the conduction bands $E_{+,\pm}$ (solid lines) and the valence bands $E_{-,\pm}$ (dashed lines) as a function of $k$. The nature of BLG dispersion is quadratic, unlike the linear dispersion for monolayer graphene; the former is due to hopping between the layers, $t_\perp$. In absence of the displacement field, $\lambda=0$, the lower conduction band $E_{+,-}$ and the upper valence band $E_{-,-}$ touch at the zero energy at $k=0$ [see in Fig.~\ref{fig:bands}(iii)(a)]; however, the bands $E_{+,+}$ and $E_{-,+}$ show a gap of $\pm t_\perp$. When $\lambda$ is  present, we notice a gap opening and Mexican-hat-like shape in $E_{+,-}$ and $E_{-,-}$. The results are shown in Figs.~\ref{fig:bands}(iii)(b)-(c). Analytical evaluation of the bands $E_{\pm,-}$ provides three extremal points $k_I=0$ and $k_{II}^\pm = \pm \sqrt{\frac{4\lambda^4+2(\lambda t_\perp)^2}{4\lambda^2+t_\perp^2}}$ which correspond to energies $E_{\pm,-} = \pm\lambda$ and $E_{\pm,-} = \pm\lambda t_\perp/\sqrt{4\lambda^2+t_\perp^2}$, respectively. This leads to the band gaps $2\lambda$ at $k_I$ and $2\lambda t_\perp/\sqrt{4\lambda^2+t_\perp^2}$ at $k_{II}^\pm$. Notably, when $\lambda\ll t_\perp$, the two gaps approximately become equal for $k_I$ and $k_{II}^\pm$ as can be seen in Fig.~\ref{fig:bands}(iii)(b). In contrast, they are different when $\lambda$ is comparable to $t_\perp$, i.e. $\lambda\lesssim t_\perp$, see in Fig.~\ref{fig:bands}(iii)(c). The minimum and maximum of $E_{+,-}$ are respectively $E_{\rm min}=\lambda t_\perp/\sqrt{4\lambda^2+t_\perp^2}$ and $E_{\rm max}=\lambda$.
\section{Model and Formalism}\label{sec:model-formalism}
In order to investigate the scattering processes and the transport properties, we consider an NS junction on the gapped BLG sheet formed at $x=0$ in $x$-$y$ plane and assume that $x<0$ is the region N, while $x>0$ occupies the region S. The superconductivity in the S region can be induced through the proximity effect by covering an external $s$-wave superconducting electrode~\cite{Volkov1995,Bee2006}. 
We employ the Dirac-Bogoliubov-de Gennes (DBdG) equation~\cite{Gennes1966,Bee2006} which couples electron with the time-reversed hole excitation wavefunctions via the superconducting (SC) pair potential. It reads
\begin{align}
    \begin{pmatrix}
        {\cal H}-E_F & \Delta(\r) \\[0.3em]
        \Delta(\r) & E_F-{\cal T}{\cal H}{\cal T}^{-1}
    \end{pmatrix}
    \begin{pmatrix}
        u_e \\[0.3em]
        v_h
    \end{pmatrix}  = \veps 
    \begin{pmatrix}
        u_e \\[0.3em]
        v_h
    \end{pmatrix}
    \label{eq:DBdG}
\end{align}
where $\veps\ge0$ is the excitation energy, $u_e (v_h)$ is the electron (hole) spinor wavefunction, and ${\cal T}$ denotes time-reversal operator. We consider the SC gap to be uniform with $\Delta(\r)=\Delta\Theta(x)$ where $\r=(x,y)$ and $\Theta(x)$ is the Heaviside step function. Also, ${\cal H}$ = diag(${\cal H}_{+}, {\cal H}_{-}$) with ${\cal H}_{\pm} = H_{\pm}+ U(\r)$ where $H_{\pm}$ are given in Eq.~\eqref{eq:H0} for the valley $\eta=\pm$ and $U(\r)=-U_0\Theta(x)$ is a doping potential in S region. For $U_0\gg E_F$, the Fermi wavelength on the N side is much larger compared to the S side which enables $\Delta(\r)$ to attain its bulk value $\Delta$. In the absence of magnetic field, the Eq.~\eqref{eq:DBdG} preserves the time-reversal symmetry such that ${\cal T}{\cal H}{\cal T}^{-1} = {\cal H}$. 
However, this operation interchanges the valley $K$ to $K'$ and vice-versa. So, we can decouple the Eq.~\eqref{eq:DBdG} into two sets of eight
equations, 
\begin{align}
    \begin{pmatrix}
        {\cal H}_\eta-E_F & \Delta(\r)  \\[0.3em]
        \Delta(\r) & E_F-{\cal H}_\eta
    \end{pmatrix}
    \begin{pmatrix}
        u_e \\[0.3em]
        v_h
    \end{pmatrix}  = \veps 
    \begin{pmatrix}
        u_e \\[0.3em]
        v_h
    \end{pmatrix}.
    \label{eq:DBdG-new}
\end{align}
Notice that the subscript $\eta$ in the above equation serves another purpose; the DBdG Eq.~\eqref{eq:DBdG-new} for $\eta=+(-)$ couples electron excitation from the valley $K(K')$ to hole excitation at the valley $K'(K)$.

To find the excitation energy spectrum for Eq.~\eqref{eq:DBdG-new} on both sides of the junction, we follow the similar calculation procedures as given in Ref.~\cite{Bee2006} and consider a plane-wave solution $(u_e~v_h)^{T} e^{ik_x x+ik_y y}$. For the N side, the excitation energies for electron ($e$) and hole ($h$) are
\begin{align}
    \veps_{\nu,\pm}^e =  E_{\nu,\pm} - E_F \quad \mathrm{and}\quad \veps_{\nu,\pm}^{h}  =  E_F - E_{\nu,\pm}
    \label{eq:excitationN}
\end{align}
Similarly, for the S side, the excitation energies for electron-like and hole-like quasiparticles are
\begin{subequations}
\begin{align}
    \veps_{\pm,\pm}^{(S,1)} &=  \pm \sqrt{[(U_0+E_F) \pm \gamma_1]^2 + \Delta^2} \\
    \veps_{\pm,\pm}^{(S,2)}  &= \pm \sqrt{[(U_0+E_F) \pm \gamma_2]^2 + \Delta^2}  
\end{align}
\end{subequations}
where $\gamma_i = \sqrt{k^2 +  \frac{t_\perp^2}{2}+ \lambda^2 + (-1)^i \sqrt{\big(\frac{t_\perp^2}{2}\big)^2 + k^2 \left(4\lambda^2+t_\perp^2\right)}}$ for $i=1,2$. 

Since the system is translationally invariant along the $y$-direction, the transverse momentum $k_y$ is conserved during the scattering process. Therefore, for a given $\veps$ and $k_y$, we solve the Eq.~\eqref{eq:DBdG-new} on the N side to obtain the state vector and longitudinal momentum $k_x$. The $\eta$-dependent states for electron and hole
are
{\color{mycol}\begin{align}
    u_{e}^\eta(\veps,k_x) &= \frac{1}{N_e}
    \begin{pmatrix}
    	- t_\perp k_{-}(\lambda+\veps_+) \\[0.2em]
        t_\perp (\lambda^2-\veps_+^2) \\[0.2em]
        [(\lambda-\veps_+)^2- k^2](\lambda+\veps_+) \\[0.2em]
        k_{+}[(\lambda-\veps_+)^2 - k^2]
        \label{eq:state-elect}
    \end{pmatrix} \\
    v_h^\eta(\veps,k_x) &= \frac{1}{N_h}
    \begin{pmatrix}
    	 - t_\perp k_{-}(\lambda+\veps_{-}) \\[0.2em]
        t_\perp (\lambda^2-\veps_{-}^2) \\[0.2em]
        [(\lambda-\veps_{-})^2 - k^2](\lambda+\veps_{-}) \\[0.2em]
        k_{+}[(\lambda-\veps_{-})^2 - k^2]
        \label{eq:state-hole}
    \end{pmatrix}
\end{align}}
where $\veps_\pm=(E_F\pm\veps)$, $N_e (N_h)$ is normalization constant fulfilling the condition ${u_e^\eta}^\dagger u_e^{\eta}=1 ({v_h^\eta}^\dagger v_h^{\eta}=1)$, and  the corresponding longitudinal momenta for $e$ and $h$ are
\begin{align}
    \pm k_{x}^{e\tau} &=\pm\sqrt{\veps_{+}^2+\lambda^2+\tau\Sigma_{e}-k_y^2}\\
    \pm k_{x}^{h\tau} &=\pm\sqrt{\veps_{-}^2+\lambda^2+\tau\Sigma_{h}-k_y^2}
\end{align}
with index $\tau=\pm$ (denoting, number of incident modes) and $\Sigma_{e(h)}= \sqrt{(4\lambda^2+t_\perp^2)\veps_{+(-)}^2 - (\lambda t_\perp)^2}$. However, for S region, finding an explicit analytical expression for the state vector is a substantially difficult task as the non-zero $\Delta$ couples $u_e$ and $v_h$. Additionally, the matrix size of the DBdG Eq.~\eqref{eq:DBdG-new} for the gapped BLG is double in comparison to the monolayer graphene~\cite{Bee2006}. Therefore, the state vector in this region is calculated numerically, say, $u_S^\eta(\veps, k_x)$. Nevertheless, we deduce an analytical form for longitudinal momenta: $\{\pm k_{x,\pm}^{(S, 1)}, \pm k_{x,\pm}^{(S, 2)}\}$ with  $k_{x,\pm}^{(S, i)} = \sqrt{\Gamma_i^2 + \lambda^2 - k_y^2 \pm \sqrt{ (4\lambda^2+t_\perp^2) \Gamma_i^2 - (\lambda t_\perp)^2}}$ where $\Gamma_i = (U_0+E_F) -(-1)^i \sqrt{\veps^2-\Delta^2}$ for $i=1,2$.

Using these states, we construct the scattering wavefunctions for both sides which consist Andreev and normal reflection coefficients and transmission coefficients, and that are obtained by demanding the continuity of the wavefunctions at the junction $x=0$. By utilizing these coefficients, we calculate the Andreev and normal reflections probabilities and the differential conductance in the framework of Blonder-Tinkham-Klapwijk (BTK) formalism~\cite{Btk1982}. See the Appendix~\ref{app:append_sec1} and~\ref{app:append_sec2} for the technical details that are given for the two regimes $\lambda\ll t_\perp$ and $\lambda\lesssim t_\perp$, respectively.
\begin{figure}[htb]
	\includegraphics[width=0.47\textwidth]{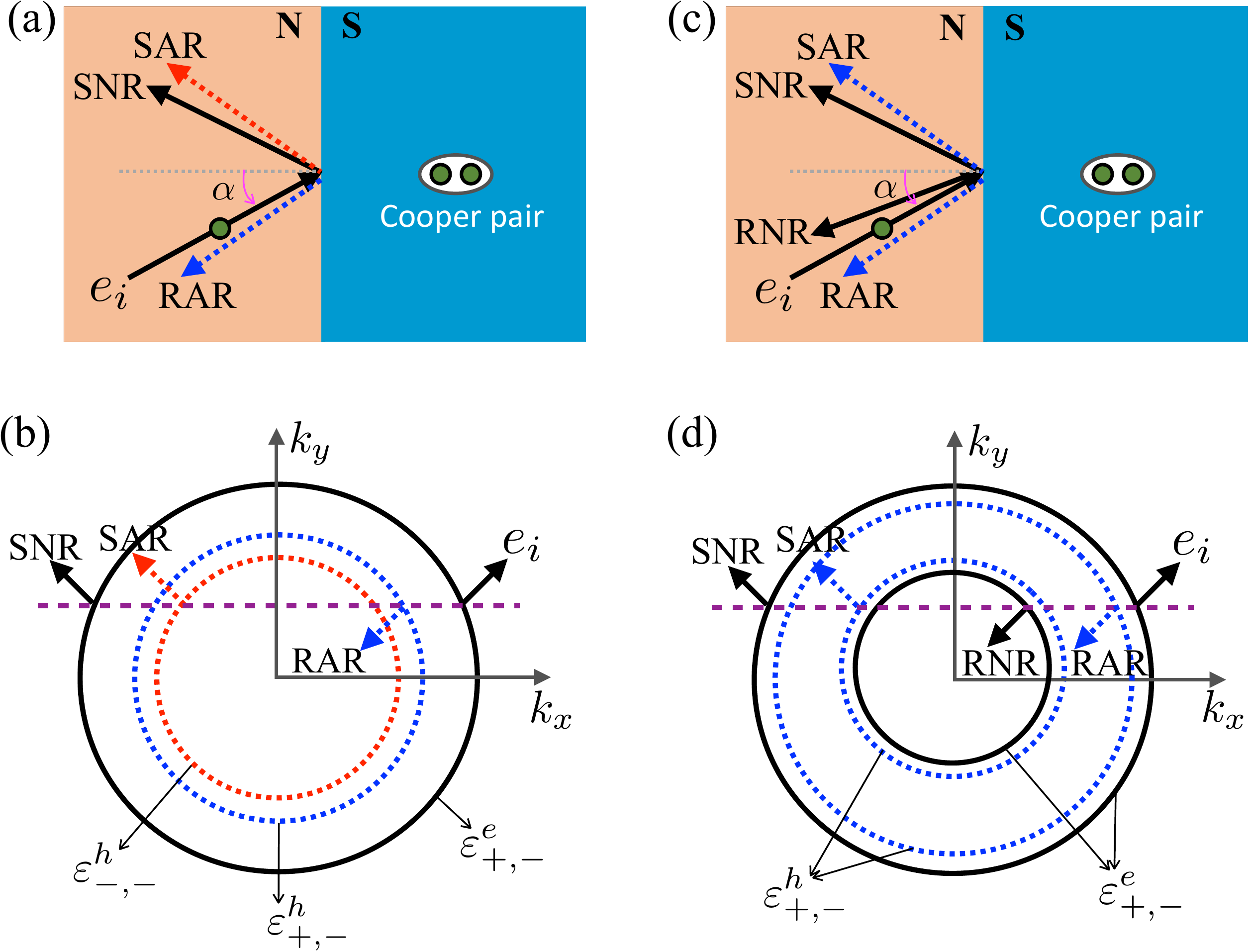}
	\caption{Schematic depiction of the gapped bilayer graphene-superconductor junction and the possible reflection processes when an electron $e_i$ (solid arrow with filled green circle) is incident at an angle $\alpha$: (a)-(b) $\lambda\ll t_\perp$ and (c)-(d) $\lambda\lesssim t_\perp$.  The other solid arrows and dotted arrows represent respectively the electron reflections and hole reflections. In (b) and (d), the solid and dotted concentric circles in $k_x$-$k_y$ plane denote isoenergy contours for electron and hole excitations, see Eq.~\eqref{eq:excitationN}, and arrows at the horizontal dashed-purple-line are the corresponding group velocities. For the incident electron $e_i$ in (a)-(b), the reflected hole is RAR when $\veps<E_F$ (dotted blue arrow) and is SAR when $\veps>E_F$ (dotted red arrow). The electron-hole conversion for RAR is intraband (only the conduction band) however SAR is due to the interband conversion (conduction band to valence band). In contrast, double Andreev reflections (RAR and SAR) and double normal reflections (RNR and SNR) exist simultaneously for (c)-(d) and the electron-hole conversion is always intraband type.}
	\label{fig:scatt-processes}
\end{figure}

In Fig.~\ref{fig:scatt-processes}, we have schematically shown the possible reflection processes involved in both regimes. {\color{mycol}For $\lambda\ll t_\perp$ regime, the Figs.~\ref{fig:scatt-processes}(a)-(b) show the usual RAR, SAR, and SNR reflections and the involved excitation energy contours for electron and hole that participate in the scattering process. As $\lambda$ is small, it  only opens a band gap $\sim 2\lambda$ [see in Fig.~\ref{fig:bands}(iii)(b)] and shows the usual reflections~\cite{Ludwig2007,Efetov2016}. However, for $\lambda\lesssim t_\perp$ regime in Figs.~\ref{fig:scatt-processes}(c)-(d), four reflections happen simultaneously which are the RAR, SAR, SNR, and RNR~\cite{Cheng2020}. The extra RNR process occurs because of the inversion of the lower conduction band [see in Fig.~\ref{fig:bands}(iii)(c)] for large $\lambda$. This supports two isoenergy-contours for electron excitation $\veps_{+,-}^e$ and two for hole excitation $\veps_{+,-}^h$ when Fermi energy $E_F$ is set in-between the maximum and minimum of the lower conduction band. Consequently, the SNR and RNR from $\veps_{+,-}^e$ and the RAR and SAR from $\veps_{+,-}^h$ occur, which are shown in Fig.~\ref{fig:scatt-processes}(d).} 

In the next section, we present the results for the reflection probabilities and differential conductance at zero temperature and discuss them in detail. They are calculated by using the Eqs.~\eqref{eq:reflect1},~\eqref{eq:cond1},~\eqref{eq:reflect2}, and~\eqref{eq:cond2}. For numerical calculation, we fix $\Delta=1$ and set all energy parameters in units of $\Delta$. Since the bulk SC pair potential can be achieved $\Delta\sim1.2$ meV (by depositing \ce{NbSe2} on the BLG sheet~\cite{Efetov2016}) and the interlayer coupling in BLG is roughly $t_\perp\sim 0.39$ eV, we present all results for $t_\perp=400\Delta$, except in Figs.~\ref{fig:Gns_lam0}(c)-(d),~\ref{fig:Gns_lamp1}(b), and~\ref{fig:GnsEf_depend}(c)-(d).
\section{Results}\label{sec:result}
\subsection{Small displacement field, $\lambda\ll t_\perp$}\label{subsec:small-lam}
In this subsection, we work in regime $(\lambda, E_F, \veps, \Delta)\ll t_\perp$ which set the longitudinal momenta for $\tau=-$ mode $k_x^{e-}$ and $k_x^{h-}$ imaginary, and consequently the corresponding state vectors in Eqs.~\eqref{eq:state-elect} and~\eqref{eq:state-hole} become evanescent type. So, we consider only $\tau=+$ mode for the incident electron which governs the scattering mechanism. The possible reflection processes are depicted in Figs.~\ref{fig:scatt-processes}(a) and~\ref{fig:scatt-processes}(b), and the reflection probabilities and differential conductance formula are given in Appendix~\ref{app:append_sec1}.
\subsubsection{Reflection probabilities}
\begin{figure}[htb]
    \includegraphics[width=0.445\textwidth]{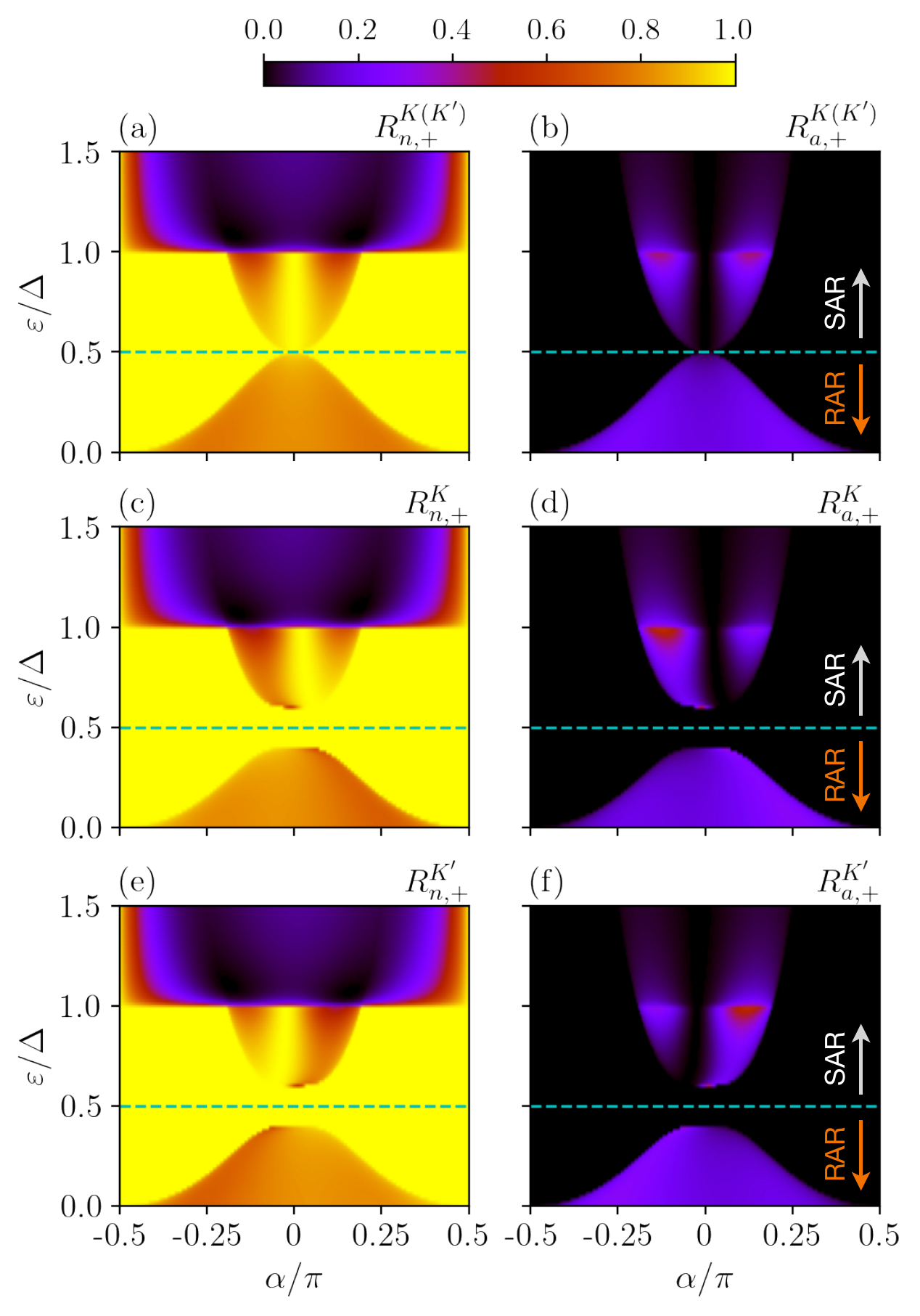}
    \caption{Heatmap plots for the normal reflection $R_{n,+}^\eta(\veps, \alpha)$ and Andreev reflection $R_{a,+}^\eta(\veps, \alpha)$ at valley $\eta\equiv K, K'$ for (a)-(b) $\lambda=0$ and (c)-(f) $\lambda=0.1\Delta$.  The other parameters are $E_F=0.5\Delta$ and $U_0=10\Delta$. The horizontal dashed-line is at $\veps=E_F$ where $R_{a,+}^\eta$ vanishes and $R_{n,+}^\eta$ becomes $1$ when $\lambda=0$, and it widens to the range $(E_F-\lambda)<\veps<(E_F+\lambda)$ of width $2\lambda$ when $\lambda\ne0$.}
    \label{fig:reflect}
\end{figure}

We present the normal and Andreev reflection probabilities, $R_{n,+}^\eta$ and $R_{a,+}^\eta$, with respect to the excitation energy $\veps$ and incident angle $\alpha$. Figure~\ref{fig:reflect} shows $R_{n,+}^\eta$ and $R_{a,+}^\eta$ plots at valley $\eta=K (K')$ for parameters $E_F=0.5\Delta$ and $U_0=10\Delta$. In Figs.~\ref{fig:reflect}(a)-(b), we set the displacement field $\lambda=0$, while it is set $\lambda=0.1\Delta$ in Figs.~\ref{fig:reflect}(c)-(f). For the excitation range $0<\veps<1.5\Delta$, the scattering process involves bands $\veps^{e}_{+,-}$, $\veps^{h}_{+,-}$, and $\veps^{h}_{-,-}$ from the N side; however, only band $\veps_{+,-}^{(S,1)}$ is involved from the S side. The incident electron always come from $\veps^{e}_{+,-}$ and the reflected hole belongs to $\veps^{h}_{+,-}$ for $0<\veps<(E_F-\lambda)$ whereas it comes from $\veps^{h}_{-,-}$ for $\veps>(E_F+\lambda)$. When $\veps>\Delta$, the transmitted quasiparticles (electron-like as well as hole-like) belong to $\veps_{+,-}^{(S,1)}$.

In Figs.~\ref{fig:reflect}(a)-(b), $R_{n,+}^{K(K')}$ and $R_{a,+}^{K(K')}$ are valley degenerated and symmetric with respect to $\alpha$. When $0<\veps<E_F$, at a given $\veps$, the $R_{n,+}^\eta$ is enhanced while $R_{a,+}^\eta$ is weakened as $\alpha$ increases from $0$ to $\pi/2$. The $R_{n,+}^\eta(R_{a,+}^\eta)$ for  $E_F<\veps<\Delta$ starts approximately equal to $1(0)$ [see the colorbar of Fig.~\ref{fig:reflect}] at $\alpha=0$ and decreases (increases) with $\alpha/\pi\lesssim0.25$ and again reaches to $1(0)$ for further increasing. We notice that $R_{n,+}^\eta+R_{a,+}^\eta=1$ as no transmissions are allowed in subgap ($\veps<\Delta$) region, while it is weakened $R_{n,+}^\eta+R_{a,+}^\eta \ne 1$ in $\veps>\Delta$ region because the quasiparticles transmission also happen. When $\lambda=0.1\Delta$ in Figs.~\ref{fig:reflect}(c)-(f), both $R_{n,+}^\eta$ and $R_{a,+}^\eta$ follow similar behaviour as in Figs.~\ref{fig:reflect}(a)-(b), but they are now slightly asymmetric about $\alpha$ for each valley $\eta=K, K'$. Changing the valley from $K\to K'$, this asymmetry is reversed, i.e. $R_{n,+}^{K}(\alpha)=R_{n,+}^{K'}(-\alpha)$ and $R_{a,+}^{K}(\alpha)=R_{a,+}^{K'}(-\alpha)$, which is a direct consequence of the layer asymmetry as $\lambda$ is nonzero. The presence of $\lambda$ opens a gap $\sim 2\lambda$ at $\veps=E_F$ (cyan dashed-line) for the incident angle $\alpha=0$ as neither $\veps^{h}_{+,-}$ nor $\veps^{h}_{-,-}$ is available for the excitation range $(E_F-\lambda)<\veps<(E_F+\lambda)$, and as a result, Andreev reflection vanishes $R_{a,+}^\eta=0$ and normal reflection reaches to $R_{n,+}^\eta=1$.
\subsubsection{Differential conductance}
\begin{figure}[htb]
	\includegraphics[width=0.475\textwidth]{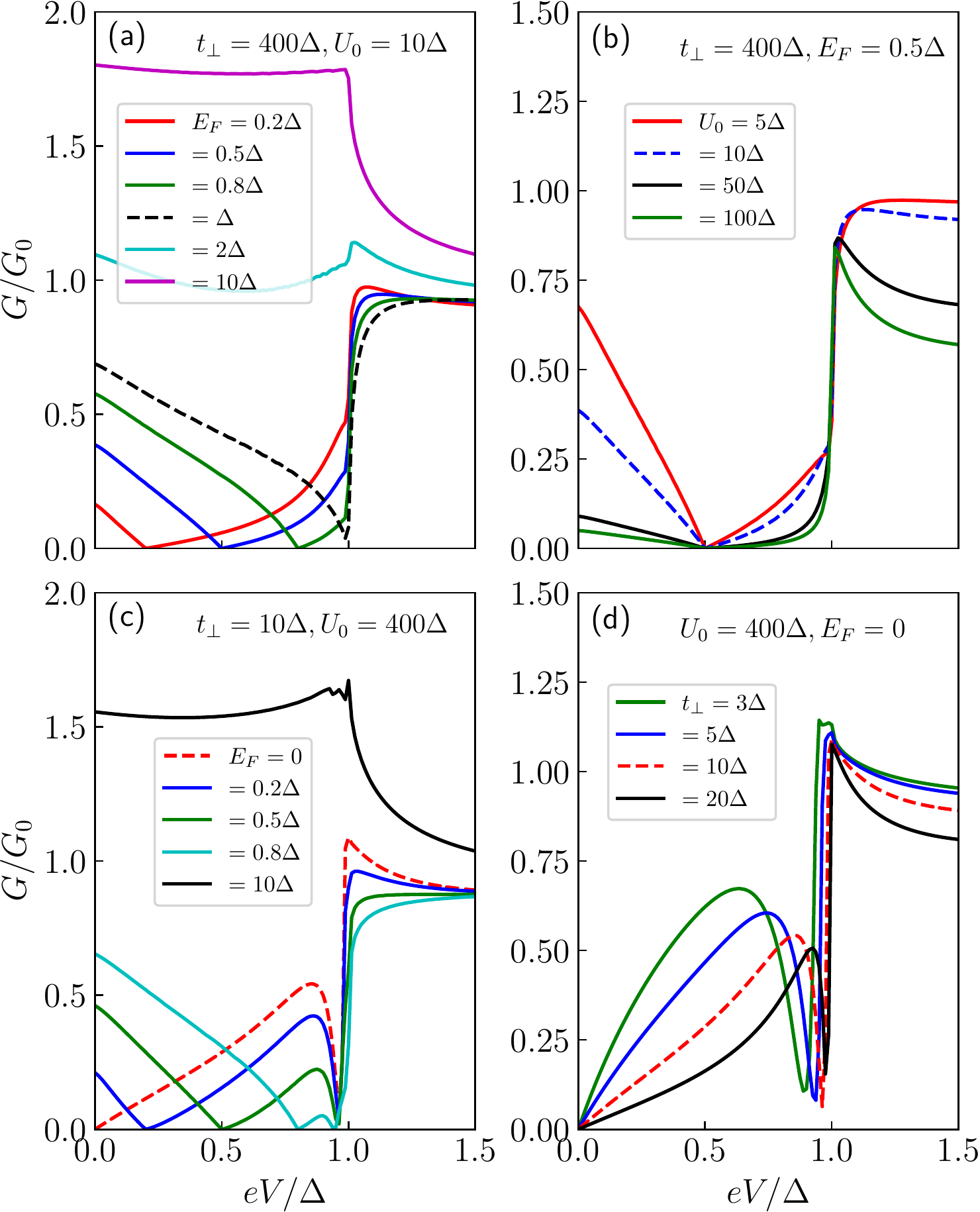}
	\caption{Normalized conductance $G/G_0$ vs. $eV/\Delta$, calculated from Eq.~\eqref{eq:cond1}, in the absence of displacement field $\lambda=0$. (a) The $G/G_0$ for several fixed $E_F$ in limit $t_\perp\gg U_0$. (b) The effect of $U_0$ on $G/G_0$ for $E_F=0.5\Delta$. (c) In opposite limit $t_\perp\ll U_0$, a dip in $G/G_0$ is observed near the gap $eV\sim\Delta$ when $E_F<\Delta$. (d) For $E_F=0$, the dip near the SC gap moves closure to $\Delta$ as the interlayer coupling $t_\perp$ increases.}
	\label{fig:Gns_lam0}
\end{figure}

Figure~\ref{fig:Gns_lam0} shows the normalized conductance $G/G_0$ versus applied bias voltage $\veps=eV$ when $\lambda=0$ which is obtained by using Eq.~\eqref{eq:cond1}. First, we discuss the result in the limit $t_\perp\gg U_0$ in Figs.~\ref{fig:Gns_lam0}(a)-(b) for $t_\perp=400\Delta$ and $U_0=10\Delta$. This limit corresponds to the work in Ref.~\cite{Efetov_PRB2016} where authors calculated the subgap conductance approximately. As discussed in the previous subsection, the scattering process always involves $\veps^{e}_{+,-}$ for electron, and $\veps^{h}_{+,-}$ for the reflected hole when $eV<E_F$ (RAR) else $\veps^{h}_{-,-}$ when $eV>E_F$ (SAR), whereas $\veps_{+,-}^{(S,1)}$ accounts for the quasiparticles transmission when $eV>\Delta$.

In Fig.~\ref{fig:Gns_lam0}(a), we plot $G/G_0$ for several fixed $E_F$. For $E_F<\Delta$, it begins with a finite value and starts decreasing as $eV$ increases and vanishes at $eV=E_F$ since no Andreev reflection happens at any incident angle. On further increasing $eV$, it again rises, exhibiting a singularity at $eV=\Delta$ similar to the ordinary NS junction~\cite{Btk1982}. The $G/G_0$ becomes weak when $eV>\Delta$, due to the quasiparticles transmission. However, for $E_F\geq\Delta$, only RAR contributes to the subgap conductance and reaches $G/G_0\to2$ for large $E_F$, see the curve for $E_F=10\Delta$. The Fermi energy mismatch at the junction, due to the finite doping potential $U_0 =10\Delta$, prevents achieving the maximum value of $G/G_0=2$. In Fig.~\ref{fig:Gns_lam0}(b), we show $U_0$-dependent $G/G_0$ at $E_F=0.5\Delta$. It shows a strong suppression on increasing $U_0$ as the Fermi energy mismatch at the junction increases with $U_0$ which reduces the AR, leading to the suppression. Our results are numerically exact and are in agreement with the findings in Ref.~\cite{Efetov_PRB2016}.

Next, we set the parameters $t_\perp=10\Delta$ and $U_0=400\Delta$ and work in the opposite limit $t_\perp\ll U_0$ to observe the dip in $G/G_0$ near the gap $eV\sim\Delta$ which was attributed to the pseudospin-$1$ effect in Ref.~\cite{Ludwig2007}. Similar to the case in Figs.~\ref{fig:Gns_lam0}(a)-(b), the scattering process for $t_\perp\ll U_0$ involves the bands $\veps^{e}_{+,-}$ and $\veps^{h}_{+,-}$ ($\veps^{h}_{-,-}$) for electron and reflected hole, but the quasiparticles now participate from bands $\veps_{+,-}^{(S,1)}$ and $\veps_{+,-}^{(S,2)}$ for the transmission. Consequently, we observe a dip in the conductance near $\Delta$, see in Figs.~\ref{fig:Gns_lam0}(c)-(d). The $G/G_0$ plots for $E_F=0$, $0.5\Delta$, and $10\Delta$ in Fig.~\ref{fig:Gns_lam0}(c) and for $t_\perp=3\Delta$ and $10\Delta$ in Fig.~\ref{fig:Gns_lam0}(d) are generated for the same values of parameters taken in~\cite{Ludwig2007}. We obtain exactly the same results for $G/G_0$ as obtained in Ref.~\cite{Ludwig2007}. This shows that our numerical method for the bilayer graphene-superconductor junction produces exact results, and also works in the both limits $t_\perp\gg U_0$ and $t_\perp\ll U_0$.
\begin{figure}[htb]
	\includegraphics[width=0.445\textwidth]{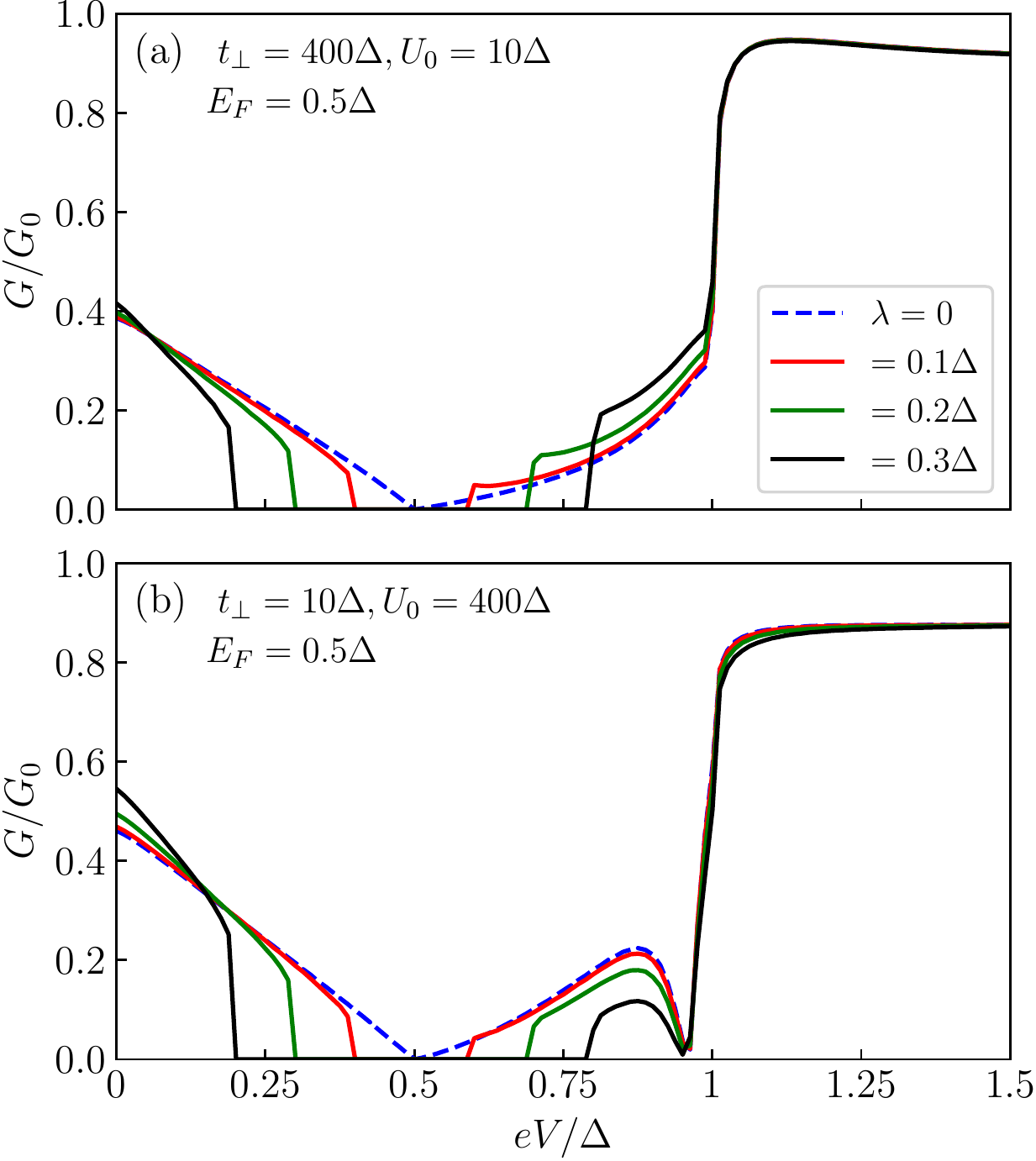}
	\caption{Normalized conductance $G/G_0$ with respect to $eV/\Delta$ at fixed $E_F$ for different $\lambda$. The top panel (a) is  for the limit $t_\perp\gg U_0$ whereas the bottom panel (b) corresponds to the limit $t_\perp\ll U_0$. The $G/G_0$ vanishes for the bias voltage range $(E_F-\lambda)<eV<(E_F+\lambda)$ with gap width $2\lambda$ and shows a strong modification in the presence of displacement field.}
	\label{fig:Gns_lamp1}
\end{figure}

Now, we present the normalized conductance $G/G_0$ behaviour in the presence of displacement field $\lambda$. The results are plotted in Fig.~\ref{fig:Gns_lamp1}(a) for the limit $t_\perp\gg U_0$ and in Fig.~\ref{fig:Gns_lamp1}(b) for the limit $t_\perp\ll U_0$. They are shown for different $\lambda$ at $E_F=0.5\Delta$. We restrict $\lambda\leq E_F$. As can be seen that the $G/G_0$ vanishes around $E_F$ for the bias range $(E_F-\lambda)<eV<(E_F+\lambda)$ at any finite $\lambda$, creating a gap of width $\sim2\lambda$ (solid lines) along the applied bias voltage axis. It happens because no Andreev reflections (neither RAR nor SAR) take place in this range as both the hole bands $\veps^{h}_{+,-}$ and $\veps^{h}_{-,-}$ are absent. This strong modification due to the finite displacement field $\lambda$ suggests that we can tune the subgap conductance and amplify the retro-to-specular crossover which would {\color{mycol}possibly} help to realise the crossover boundary experimentally~\cite{Efetov2016}, {\color{mycol}since the crossover-boundary becomes wider.}
\subsection{Large displacement field, $\lambda\lesssim t_\perp$}\label{subsec:large-lam}
This section is devoted to studying the transport properties at the junction in large displacement field regime, wherein the band structure illustrated in Fig.~\ref{fig:bands}(iii)(c) participates in the scattering process. In this regime, both the $\tau=\pm$ modes with incident electron longitudinal wave-vectors $k_x^{e\pm}$ are present. The possible reflection processes are schematically shown in Figs.~\ref{fig:scatt-processes}(c) and~\ref{fig:scatt-processes}(d). We fix $E_F$ comparable to $\lambda$, i.e. $E_F\simeq\lambda$, and set the parameters $t_\perp=400\Delta$, $E_F=80\Delta$, and $U_0=10\Delta$ for the calculation. We first calculate the reflection probabilities at a given $\lambda$ with respect to $\alpha$ and $\veps$. Using these probabilities, the conductance is calculated by integrating over $\alpha$, see the formulation and technical details in Appendix~\ref{app:append_sec2}. For brevity, we only present the results for differential conductance and discuss the key findings (the $\alpha$-dependent and $\veps$-dependent reflection probabilities are presented in Appendix~\ref{app:append_sec2} for completeness).

\begin{figure}[htb]
\includegraphics[width=0.46\textwidth]{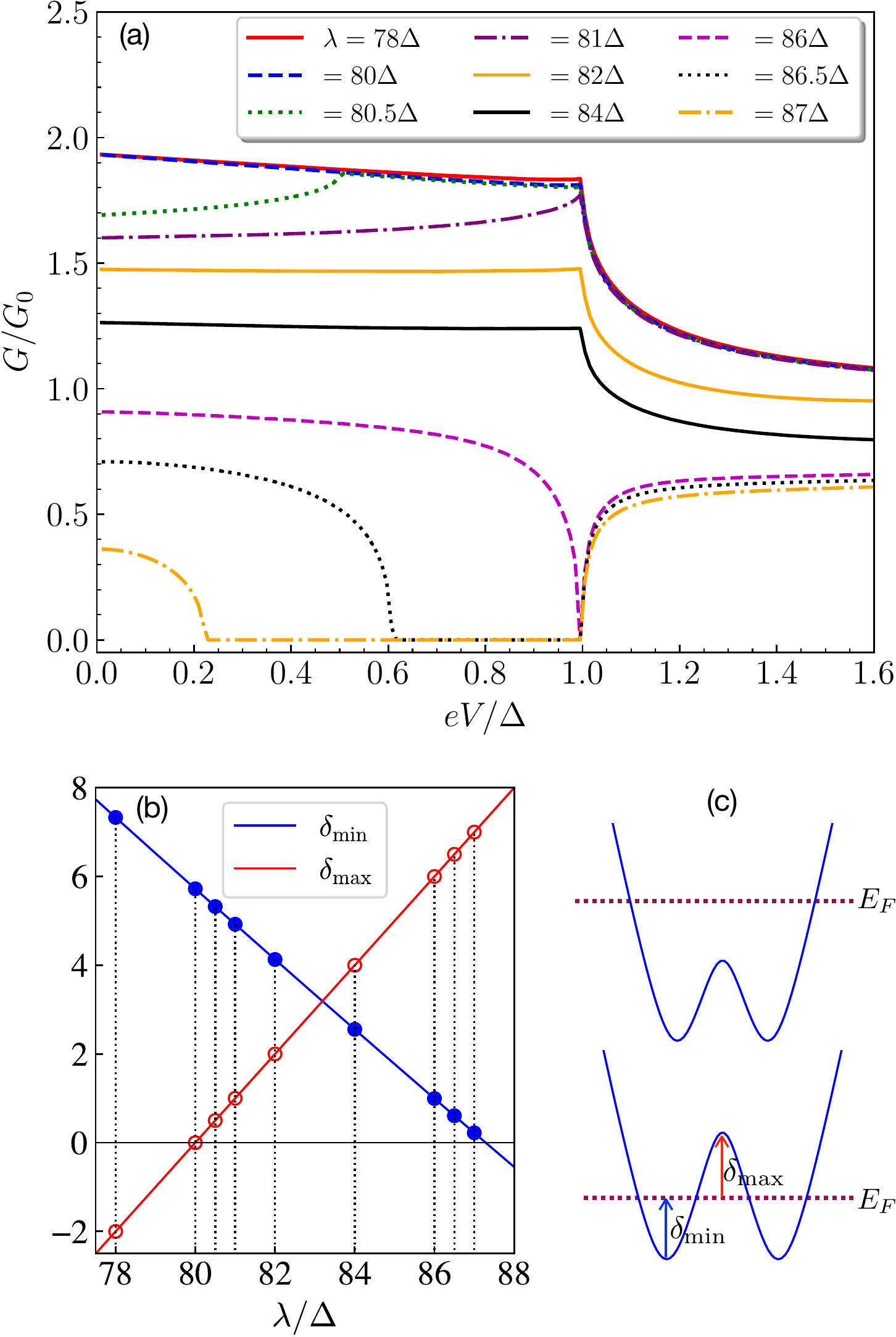}
\caption{(a) Normalized conductance $G/G_0$, calculated by using the Eq.~\eqref{eq:cond2}, with $eV/\Delta$ for several $\lambda$'s. (b) Variation of $\delta_{\rm min}=E_F-E_{\rm min}$ and $\delta_{\rm max}=E_{\rm max}-E_F$ with $\lambda/\Delta$. The filled and empty circles correspond to the points $(\delta_{\rm min}/\Delta,\delta_{\rm max}/\Delta)$ = $(7.33, -2)$, $(5.72, 0)$, $(5.32,0.5)$, $(4.92,1)$, $(4.13,2)$, $(2.55,4)$, $(1,6)$, $(0.61,6.5)$, $(0.22,7)$ for the selected $\lambda$'s in (a). The schematic plot of the conduction band in (c) is shown for the situations when $\lambda<E_F$ (top panel) and $\lambda\geq E_F$ (bottom panel).}
\label{fig:Gns_large_lam}
\end{figure}

Figure~\ref{fig:Gns_large_lam}(a) presents the normalized conductance $G/G_0$ versus $eV/\Delta$ for several values of $\lambda$ taken around $E_F$. The scattering process involves only the conduction band excitations $\veps^{e}_{+,-}$ and $\veps^{h}_{+,-}$ for electron and hole, respectively. So, the electron-hole conversion for the Andreev reflections is always intraband in nature. To better understand the behavior of $G/G_0$, we also plot quantities $\delta_{\rm min}=E_F-E_{\rm min}$ and $\delta_{\rm max}=E_{\rm max}-E_F$ with respect to $\lambda$ in Fig.~\ref{fig:Gns_large_lam}(b), which are depicted in the bottom panel of Fig.~\ref{fig:Gns_large_lam}(c). The filled and open circles, in Fig.~\ref{fig:Gns_large_lam}(b), are at the selected $\lambda$ points for which $G/G_0$ is plotted. The top panel of Fig.~\ref{fig:Gns_large_lam}(c) shows the situation when $E_F$ lies above $E_{\rm max}$ in the conduction band, i.e, $E_F>E_{\rm max}$.

For $\lambda=78\Delta$, in Fig.~\ref{fig:Gns_large_lam}(a), only the $\tau=+$ mode wave-vectors are available for electron and hole excitations for bias $0<eV<1.6\Delta$, so the SAR and RNR vanish as the $\tau=-$ mode is absent due to the condition $E_F>E_{\rm max}$. Therefore, only the RAR and SNR contribute to the conductance. $G/G_0\to2$ in the subgap region remains almost constant because $E_F\gg\Delta$ and provides a similar result as shown in Fig.~\ref{fig:Gns_lam0}(a) for $E_F=10\Delta$ and behaves as an ordinary NS junction~\cite{Btk1982}. When $\lambda=80\Delta(=E_F)$, the scattering process is still governed by $\tau=+$ mode wave-vectors for electron excitation and thus similar $G/G_0$ behavior, but now both Andreev reflections (RAR and SAR) exist along with SNR  due to the presence of $\tau=\pm$ wave-vectors for hole excitation. Next, we notice that the $G/G_0$ starts with a lower value and rises again to the previous value at the critical bias $(eV)_c = 0.5\Delta$ for $\lambda=80.5\Delta$ (dotted green curve, $\delta_{\rm max}=0.5\Delta$) and $(eV)_c = \Delta$ for $\lambda=81\Delta$ (dot-dashed purple curve, $\delta_{\rm max}=\Delta$) because the $\tau=\pm$ modes participate for electron and hole in range $0<eV<(eV)_c$. This leads to the appearance of a new type of normal reflection---retro normal reflection (RNR). Correspondingly, it weakens $G/G_0$ in the region $0<eV<(eV)_c$.

In contrast, when $\lambda>(E_F+\Delta)$, both $\tau=\pm$ modes wave-vectors from $\veps^{e}_{+,-}$ are always available for electron excitation, but depending on the strength of $\lambda$ the hole excitation $\veps^{h}_{+,-}$ does not always participate in the scattering process for the whole range of $eV$. For instance, $\veps^{h}_{+,-}$ is always involved at $\lambda=82\Delta$ and $84\Delta$, and thus all four reflections contribute. The further decrease in $G/G_0$ within the subgap region is due to the presence of RNR along with SNR, whereas the quasiparticle transmissions are responsible for the decrease in the $eV>\Delta$ region. However, the hole excitation $\veps^{h}_{+,-}$ is absent when bias voltage exceeds the critical bias $(eV)_c=\Delta$, $0.61\Delta$, and $0.22\Delta$ for the displacement fields $\lambda=86\Delta$ ($\delta_{\rm min}=\Delta$), $\lambda=86.5\Delta$ ($\delta_{\rm min}=0.61\Delta$), and $\lambda=87\Delta$ ($\delta_{\rm min}=0.22\Delta$), respectively. As a result, the double Andreev reflections SAR and RAR become zero, and the non-zero double normal reflections SNR and RNR cause the zero conductance inside the gap for $(eV)_c<eV<\Delta$. These results are consistent with the discussion of the reflection probabilities, given in Appendix~\ref{app:append_sec2}. Overall, we observe that $G/G_0$ clearly shows a distinct feature which could be experimentally verified by varying the displacement field close to the Fermi energy of the normal side.
{\color{mycol}
\subsection{Experimentally-feasible results: $\lambda=0$ (in S region)}\label{subsec:exp}
In the preceding two subsections, we examined the conductance characteristics in the small and large displacement field regimes while keeping $\lambda$ non-zero and equal on both the N and S sides of the junction. Nevertheless, implementing a non-zero displacement field on the S side in an experimental setup would likely pose considerable challenges, if not impossibilities. Therefore, in this subsection, we focus on calculating the conductance in the small and large displacement field regimes when $\lambda$ is turned off on the S side; however, it is still applied on the N side.
\begin{figure}[htb]
	\includegraphics[width=0.45\textwidth]{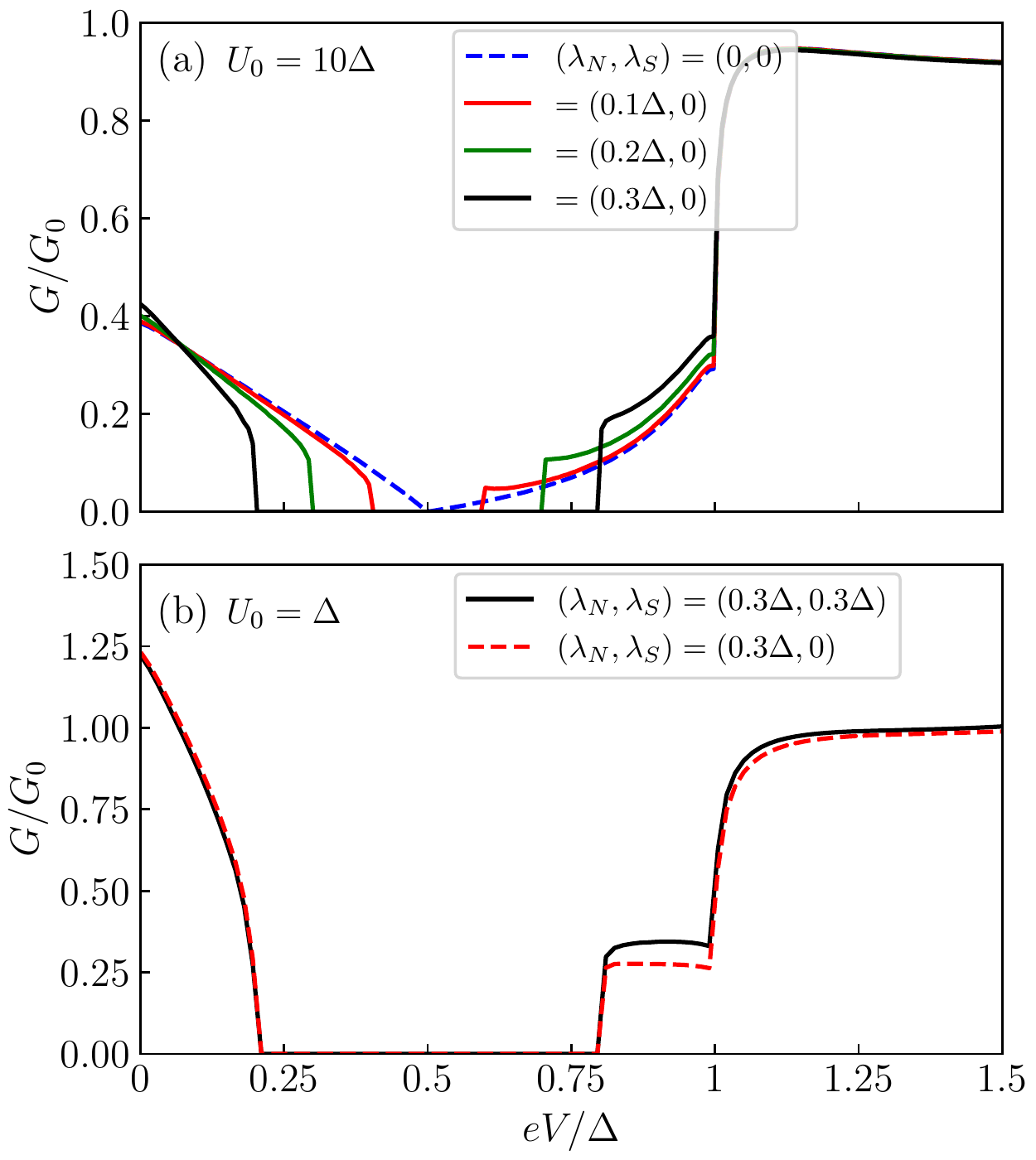}
	\caption{{\color{mycol}The normalized conductance $G/G_0$ vs. $eV/\Delta$ for different non-zero $\lambda$'s in normal region. (a) Conductance behaviour is almost exactly same as obtained in the Fig.~\ref{fig:Gns_lamp1}(a) for $U_0=10\Delta$. However, it deviates slightly in (b) for $U_0=\Delta$. The other parameters are $t_\perp=400\Delta$ and $E_F=0.5\Delta$. The subscript in $\lambda_{N(S)}$ is added to specify the displacement field $\lambda$ in N (S) region.}}
	\label{fig:Gns_lam1_exp}
\end{figure}

In Fig.~\ref{fig:Gns_lam1_exp}, we show the normalized conductance $G/G_0$ with respect to $eV/\Delta$ for different values of $\lambda$ in the regime $\lambda\ll t_\perp$. The other fixed parameters are given in the figure panels and caption. Notice that the conductance curves in Fig.~\ref{fig:Gns_lam1_exp}(a) exhibit almost exactly the same result as in Fig.~\ref{fig:Gns_lamp1}(a) even though $\lambda=0$ in the S region. This happens mainly because the Fermi energy mismatch at the junction does not change significantly since $\lambda\ll U_0$ as $U_0=10\Delta$ in the both calculation. In contrast, this difference is visible when we set $U_0=\Delta$ in Fig.~\ref{fig:Gns_lam1_exp}(a) as $\lambda\sim U_0$. However, the qualitative behaviour is still similar.

Now, we present the $G/G_0$ behaviour in the large displacement field limit, see Fig.~\ref{fig:Gns_large_lam_exp}, for the same set of parameters values as in Fig.~\ref{fig:Gns_large_lam}, except here the displacement field $\lambda$ is non-zero only on N side, i.e., $\lambda_N=\lambda$ and $\lambda_S=0$. We see that the qualitative characteristics of $G/G_0$ is similar to the Fig.~\ref{fig:Gns_large_lam}(a), but now the subgap conductance has dropped significantly as the Fermi-energy mismatch is enlarged because of $\lambda_S=0$. Consequently, the Andreev reflections (normal reflections) contribution to the conductance would be decreased (increased), reducing the subgap conductance.
\begin{figure}[htb]
	\includegraphics[width=0.475\textwidth]{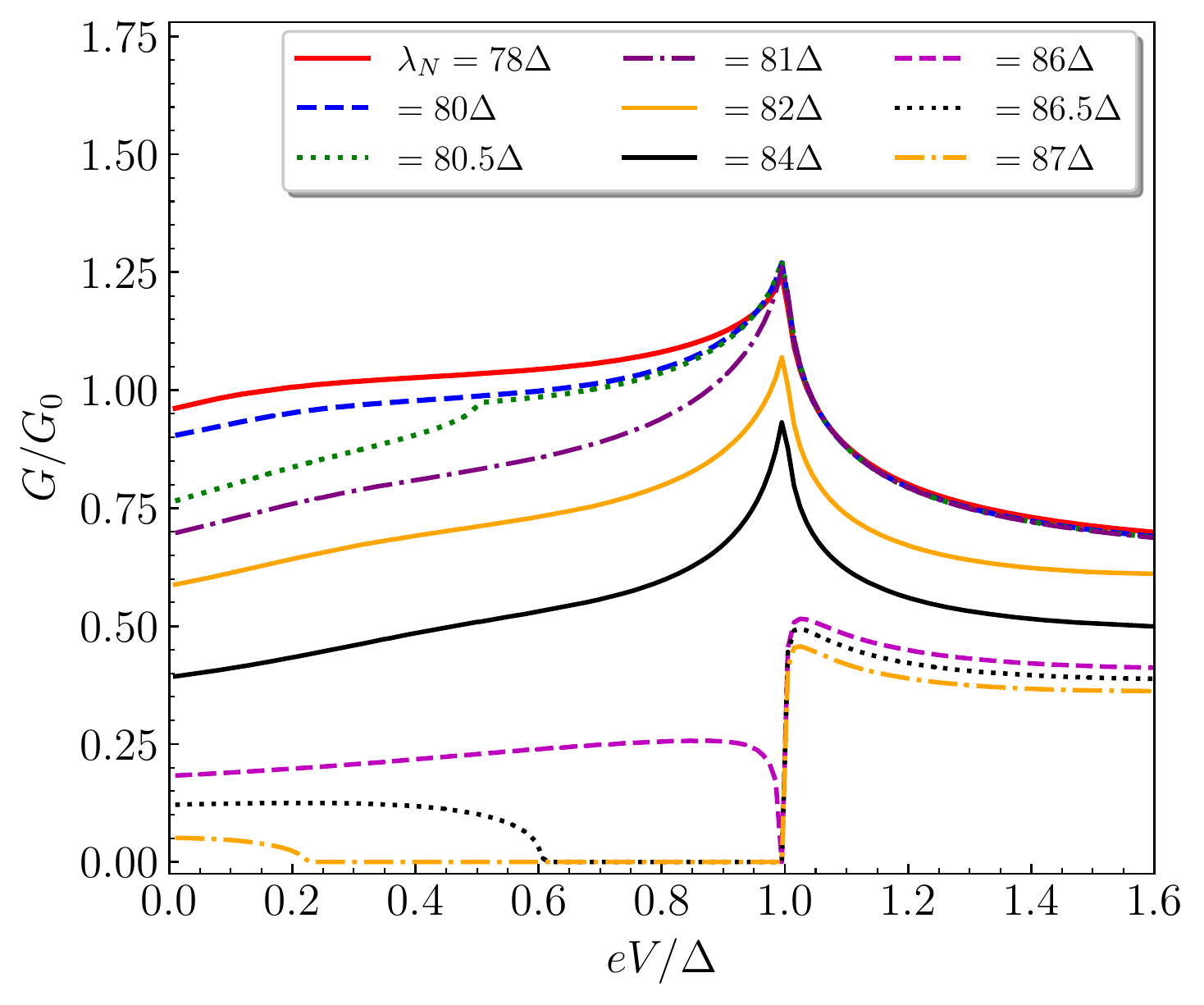}
	\caption{{\color{mycol}Normalized conductance $G/G_0$ vs. $eV/\Delta$, in large displacement field regime $\lambda\lesssim t_\perp$, for several $\lambda_N$ values while $\lambda_S=0$. The other parameters are $t_\perp=400\Delta$, $E_F=80\Delta$, and $U_0=10\Delta$.}}
	\label{fig:Gns_large_lam_exp}
\end{figure}
}
\section{Conclusion}\label{sec:conclusion}
We have studied the scattering reflection processes at the gapped bilayer graphene-superconducting junction by employing the DBdG equation within the scattering theory approach. Since $\lambda$ provides an external tool to tune the BLG band structure, we have thoroughly investigated its effect on the normal reflections, the Andreev reflections, and the experimentally accessible quantity the differential conductance in two regimes: $\lambda\ll t_\perp$ and $\lambda\lesssim t_\perp$. In the former regime, we have revisited the $\lambda=0$ case first and obtained the expected specular (retro) Andreev reflection for the excitation energy $\veps>E_F$ ($\veps<E_F$); the retro-to-specular crossing point at bias $eV=E_F$ in the subgap conductance~\cite{Ludwig2007, Efetov_PRB2016}. As our numerical calculation provides the exact result, we also observe the dip in conductance near $eV\sim\Delta$ when $t_\perp\ll U_0$, which was attributed to the pseudospin-$1$ feature in Ref.~\cite{Ludwig2007}. However, this dip does not appear when $t_\perp\gg U_0$, as is also the case in Ref.~\cite{Efetov_PRB2016}. Furthermore, introducing a non-zero small $\lambda$ broadens the retro-to-specular crossing for the bias range $|E_F-\lambda|<eV<|E_F+\lambda|$ in the subgap conductance which is a direct manifestation of the band gap between conduction and valence bands due to finite displacement field. This suggests that tuning $\lambda$ the crossover region can be easily modified and possibly obtained experimentally as the authors in~\cite{Efetov2016} struggle to observe it.

However, for the latter regime $\lambda\lesssim t_\perp$, we have shown that apart from SNR, SAR, and RAR, there exists also RNR due to the presence of the Mexican-hat-shape band structure in the gapped BLG. This is a new finding in a bilayer graphene NS junction which has never been explored as per our knowledge. The normalized conductance shows a very distinct characteristic feature when all four reflections contribute, see Fig.~\ref{fig:Gns_large_lam}, and it can be distinguished from the result when the Mexican-hat structure does not affect the scattering process.
\begin{acknowledgments}
This work was funded by the Deutsche Forschungsgemeinschaft (DFG, German Research Foundation) - 467596333. This work was partly supported by the Helmholtz Association through program NACIP.
\end{acknowledgments}
\appendix
\section{Small displacement field ($\lambda\ll t_\perp$)}\label{app:append_sec1}
In this regime, we have only the $\tau=+$ mode with the incident electron wave-vector $(k_x^{e+}, k_y^+)$; therefore, the scattering wavefunctions can be expressed as $\Psi^\eta(x)e^{ik^+_yy}$ where
\begin{align}
    \Psi^\eta(x)=& \big[u_e^\eta(\veps, k_x^{e+})e^{ik_x^{e+} x}
    + r_{n,+}^\eta u_e^\eta(\veps, -k_x^{e+})e^{-ik_x^{e+} x} \nonumber\\
    &~+ r_{n,-}^\eta u_e^\eta(\veps, -i\kappa_x^{e-})e^{\kappa_x^{e-} x} \big]\begin{pmatrix}
        1 \\[0.05em]
        0\end{pmatrix} \nonumber\\
    &~+ \big[r_{a,+}^\eta v_h^\eta(\veps,\beta k_x^{h+})e^{i\beta k_x^{h+}x} \nonumber\\
    &~+ r_{a,-}^\eta v_h^\eta(\veps, -i\kappa_x^{h-})e^{\kappa_x^{h-} x} \big]\begin{pmatrix}
        0 \\[0.05em]
        1\end{pmatrix}, \quad \mathrm{for}~x\le 0 \nonumber\\
    =&\sum_{j=1}^{4} t_j^\eta u_S^\eta(\veps, k_{x,j}^S)e^{ik_{x,j}^S x}, \quad \mathrm{for}~x\ge 0 
    \label{eq:boundary-NS1}
\end{align}
Here, $k_x^{e-} (k_x^{h-}) = -i\kappa_x^{e-}(-i\kappa_x^{h-})$ and $k_{x,j}^S$ for $j=1, ..., 4$, are the wave-vectors of side S and are chosen appropriately from $\{\pm k_{x,\pm}^{(S,1)}, \pm k_{x,\pm}^{(S, 2)} \}$. The coefficients $r_{n,\pm}^\eta$, $r_{a,\pm}^\eta$, and $t_{j=1,..,4}$ are normal reflections, Andreev reflections, and transmissions, respectively. $r_{n,-}^\eta$ and $r_{a,-}^\eta$ are always zero because of evanescent solutions for the $\tau=-$ mode. It is worth mentioning that $r_{n,+}^\eta$ is always SNR, but $r_{a,+}^\eta$ is RAR for $\veps<(E_F-\lambda)$, $\beta=1$ and SAR for $\veps>(E_F+\lambda)$, $\beta=-1$. By using the continuity condition, $\Psi^\eta|_{x=0^-}=\Psi^\eta|_{x=0^+}$, we obtain these coefficients for a given $\veps$ and incident angle $\alpha$ defined from $k^+_y =  \sqrt{(E_F+\veps)^2+\lambda^2+\Sigma_e}\sin\alpha$ where $\alpha\in[-\pi/2,\pi/2]$. So, the reflection probabilities are obtained as:
\begin{align}
    R_{n,+}^\eta&=|r_{n,+}^\eta(\veps,\alpha)|^2,~~ R_{a,+}^\eta=\frac{V_{h+}^\eta}{V_{e+}^\eta} |r_{a,+}^\eta(\veps,\alpha)|^2
    \label{eq:reflect1}
\end{align}
with velocities $V_{e+}^\eta = \langle u_e^\eta (\veps, k_x^{e+})| \frac{\partial H_\eta }{\partial k_x} |u_e^\eta (\veps, k_x^{e+})\rangle$ and  $V_{h+}^\eta = \langle v_h^\eta (\veps, k_x^{h+})| \frac{\partial H_\eta }{\partial k_x} |v_h^\eta (\veps, k_x^{h+})\rangle$. The differential conductance within the BTK framework~\cite{Btk1982} at zero temperature can be expressed as follows:
\begin{align}
    G(\veps)=\sum_{\eta}g_0(\veps)\int_{-\frac{\pi}{2}}^{\frac{\pi}{2}}\big(1 - R_{n,+}^\eta + R_{a,+}^\eta \big)\cos\alpha~d\alpha
    \label{eq:cond1}
\end{align}
with $\veps=eV$ and $g_0(\veps)=\frac{2e^2W}{h\pi}\sqrt{(E_F+\veps)^2+\lambda^2+\Sigma_e}$ where $W$ is the width of BLG sheet and factor $2$ accounts for the spin degeneracy. Also, the differential conductance for a normal-to-normal BLG junction is {\color{mycol}$G_0(\veps)=4 g_0(\veps)$ where the factor 4 is due to sum of valley index $\eta=\pm$ and incident angle $\alpha\in[-\pi/2, \pi/2]$ instead of $\alpha\in[0, \pi/2]$~\cite{Bee2006,Ludwig2007,Efetov_PRB2016}. 
}
\begin{figure}[htb]
	\includegraphics[width=0.48\textwidth]{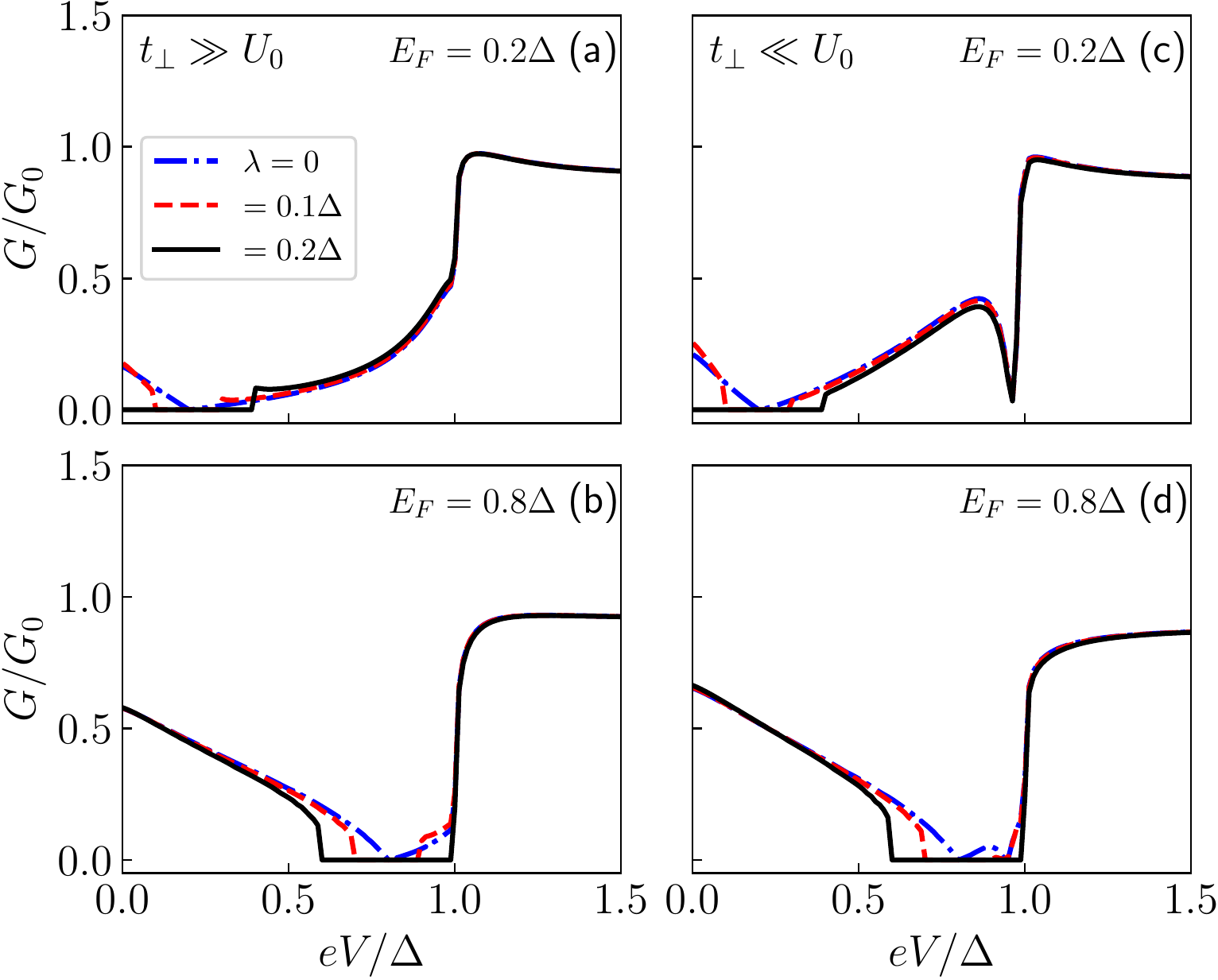}
	\caption{Behavior of the normalized conductance, $G/G_0$, with respect to $eV/\Delta$ for three values of $\lambda$ at $E_F=0.2\Delta$ and $0.8\Delta$. In (a)-(b) $t_\perp\gg U_0$ for $t_\perp=400\Delta$ and $U_0=10\Delta$ whereas in (c)-(d) $t_\perp\ll U_0$ for $t_\perp=10\Delta$ and $U_0=400\Delta$.}
	\label{fig:GnsEf_depend}
\end{figure}
\section{Effect of $\lambda$ on $G/G_0$ at different $E_F$}
In order to be consistent with the main text in subsection~\ref{subsec:small-lam}, we illustrate the effect of the displacement field $\lambda$ on the normalized conductance $G/G_0$ for two values of Fermi energy $E_F=0.2\Delta$ and $0.8\Delta$ to study the contribution of RAR and SAR in subgap region $eV<\Delta$. The results are shown in Figs.~\ref{fig:GnsEf_depend}(a)-(b) for the limit $t_\perp\gg U_0$ and in Figs.~\ref{fig:GnsEf_depend}(c)-(d) for the limit $t_\perp\ll U_0$. In both limits, when Fermi energy is set at $E_F=0.2\Delta$, the contribution of the RAR and SAR equally reduced about $E_F$ and eventually only SAR contributes to $G/G_0$ as $\lambda$ is increased to the value $\lambda=E_F$, see in Figs.~\ref{fig:GnsEf_depend}(a) and~\ref{fig:GnsEf_depend}(c). While for $E_F=0.8\Delta$ in Figs.~\ref{fig:GnsEf_depend}(b) and~\ref{fig:GnsEf_depend}(d), only the RAR contributes. Hence, varying the displacement field and setting the Fermi energy appropriately, the Andreev reflection contribution to the subgap conductance can be tuned to either SAR or RAR.
\section{Large displacement field ($\lambda\lesssim t_\perp$)}\label{app:append_sec2}
In this regime, we have both the $\tau=\pm$ modes for the incident electron with the wave-vectors $(k_{x}^{e+}, k^+_y)$ and $(-k_{x}^{e-}, k^-_y)$ where $k_y^\pm=\sqrt{(E_F+\veps)^2+\lambda^2\pm\Sigma_e}\sin\alpha$. So, the scattering wavefunctions on both sides of the junction have the form $\Psi_\tau^\eta(x)e^{ik^\tau_y y}$ with
\begin{align}
    \Psi_\tau^\eta(x)=&~
    \big[ u_e^\eta(\veps, \tau k_x^{e\tau})e^{i\tau k_x^{e\tau} x} + r_{n1, \tau}^\eta u_e^\eta(\veps, -k_x^{e+})e^{-ik_x^{e+} x} \nonumber\\
    &~+ r_{n2, \tau}^\eta u_e^\eta(\veps, k_x^{e-})e^{ik_x^{e-}x} \big]\begin{pmatrix}
        1 \\[0.05em]
        0\end{pmatrix} \nonumber\\
    &~+ \big[ r_{a1, \tau}^\eta v_h^\eta(\veps,-k_x^{h-})e^{-ik_x^{h-} x} \nonumber\\
    &~+  r_{a2, \tau}^\eta v_h^\eta(\veps, k_x^{h+})e^{ik_x^{h+} x} \big]\begin{pmatrix}
        0 \\[0.05em]
        1\end{pmatrix}, \quad \mathrm{for}~x\le 0 \nonumber\\
    =&~\sum_{j=1}^{4} t_{j,\tau}^\eta u_S^\eta(\veps, k_{x,j,\tau}^S)e^{ik_{x,j,\tau}^S x}, \quad \mathrm{for}~x\ge 0  
    \label{eq:boundary-NS2}
\end{align}
where $r_{n1,\tau}^\eta, r_{n2,\tau}^\eta, r_{a1,\tau}^\eta$, and $r_{a2, \tau}^\eta$ denote reflection coefficients, namely, SNR, RNR, SAR, and RAR for $\tau=+$ mode; however, they become RNR, SNR, RAR, and SAR for $\tau=-$ mode. The $t^\eta_{j,\tau}$ for $j=1,...,4$ are transmission coefficients. Again, using the continuity condition, we obtain the reflection probabilities as:
\begin{align}
    R_{n1,\tau}^\eta &=\frac{V_{e+}^\eta}{V_{e\tau}^\eta}|r_{n1,\tau}^\eta(\veps,\alpha)|^2,~~ R_{n2, \tau}^\eta =\frac{V_{e-}^\eta}{V_{e\tau}^\eta}|r_{n2,\tau}^\eta(\veps,\alpha)|^2 \nonumber\\
    R_{a1, \tau}^\eta &=\frac{V_{h-}^\eta}{V_{e\tau}^\eta}|r_{a1,\tau}^\eta(\veps,\alpha)|^2,~~
    R_{a2, \tau}^\eta =\frac{V_{h+}^\eta}{V_{e\tau}^\eta}|r_{a2,\tau}^\eta(\veps,\alpha)|^2
    \label{eq:reflect2}
\end{align}
with velocities $V_{e\tau}^\eta= \langle u_e^\eta(\veps, \tau k_x^{e\tau})| \frac{\partial H_\eta}{\partial k_x} |u_e^\eta(\veps, \tau k_x^{e\tau})\rangle$ and $V_{h\tau}^\eta= \langle v_h^\eta(\veps, \tau k_x^{h\tau})| \frac{\partial H_\eta}{\partial k_x} |v_h^\eta(\veps, \tau k_x^{h\tau})\rangle$. The differential conductance formula in this regime becomes
\begin{align}
G(\veps)=\sum_{\eta,\tau}g_0^\tau(\veps) \int_{-\frac{\pi}{2}}^{\frac{\pi}{2}} \Big(1 &- R_{n1,             \tau}^\eta - R_{n2, \tau}^\eta+ R_{a1, \tau}^\eta \nonumber\\
     &+ R_{a2, \tau}^\eta\Big)\cos\alpha~d\alpha
\label{eq:cond2}
\end{align}
where $\veps=eV$ and $g^\tau_0(\veps)=\frac{2e^2W}{h\pi}\sqrt{(E_F+\veps)^2 +\lambda^2 +\tau\Sigma_e}$ which leads to {\color{mycol}$G_0(\veps)=4\sum_{\tau} g^\tau_0(\veps)$ for a gapped BLG normal-to-normal junction. 
}
The results for the differential conductance have been presented in the subsections.~\ref{subsec:large-lam} and~\ref{subsec:exp} in this regime. Here, we briefly discuss the reflection probabilities for the displacement field values $\lambda=84\Delta$ and $86.5\Delta$.
\begin{figure}[htb]
	\includegraphics[width=0.47\textwidth]{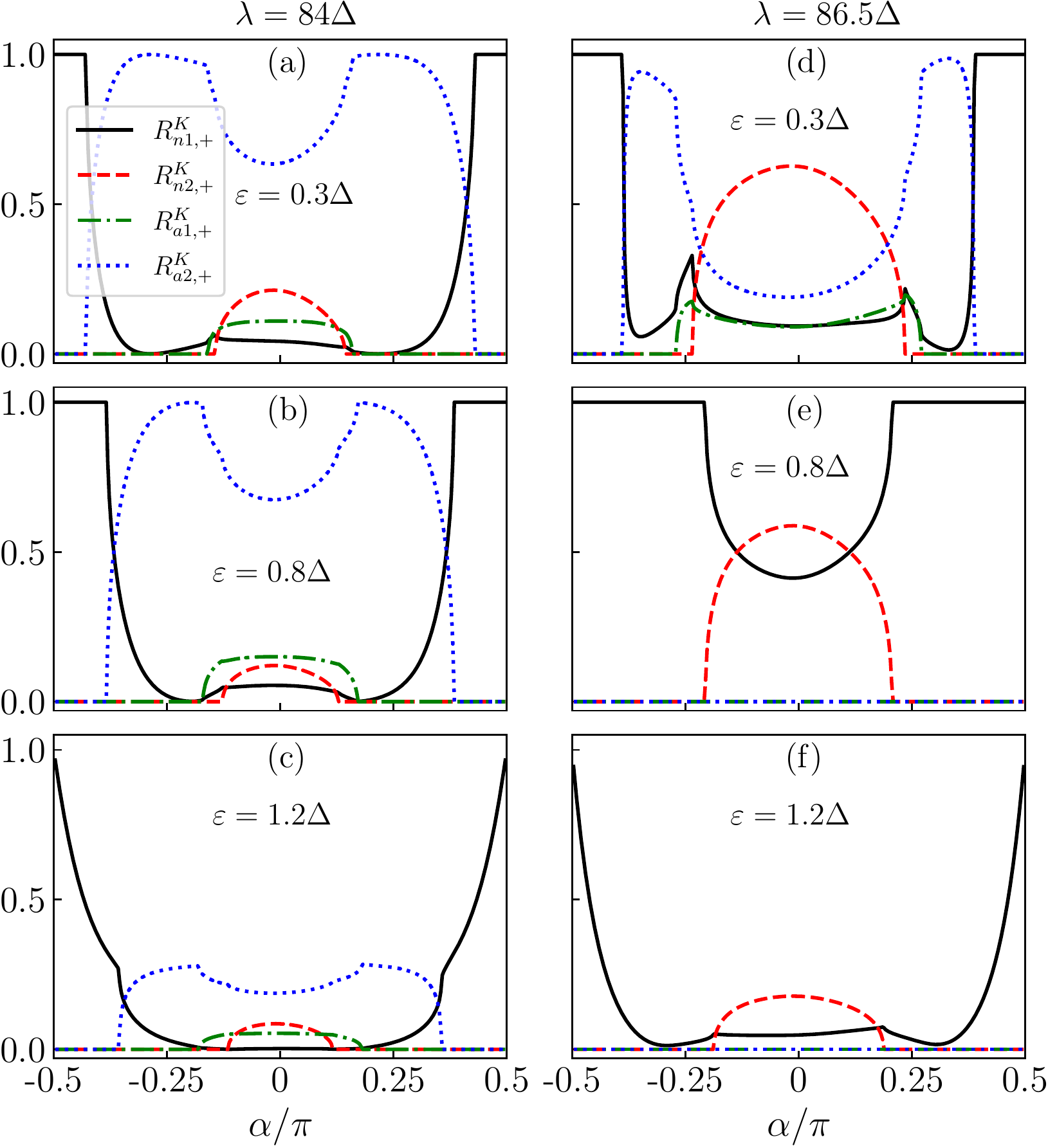}
	\caption{Reflection probabilities (at valley $K$ and $\tau=+$) vs. $\alpha$ for $E_F=80\Delta$, $t_\perp=400\Delta$, and $U_0=10\Delta$ with excitation energy $\veps=0.3\Delta$ in (a) and (d), $\veps=0.8\Delta$ in (b) and (e), and $\veps=1.2\Delta$ in (c) and (f). The displacement fields for left and right panels are $\lambda=84\Delta$ and $86.5\Delta$.}
	\label{fig:Reflam80_alpha_dep}
\end{figure}
\begin{figure}[htb]
	\includegraphics[width=0.47\textwidth]{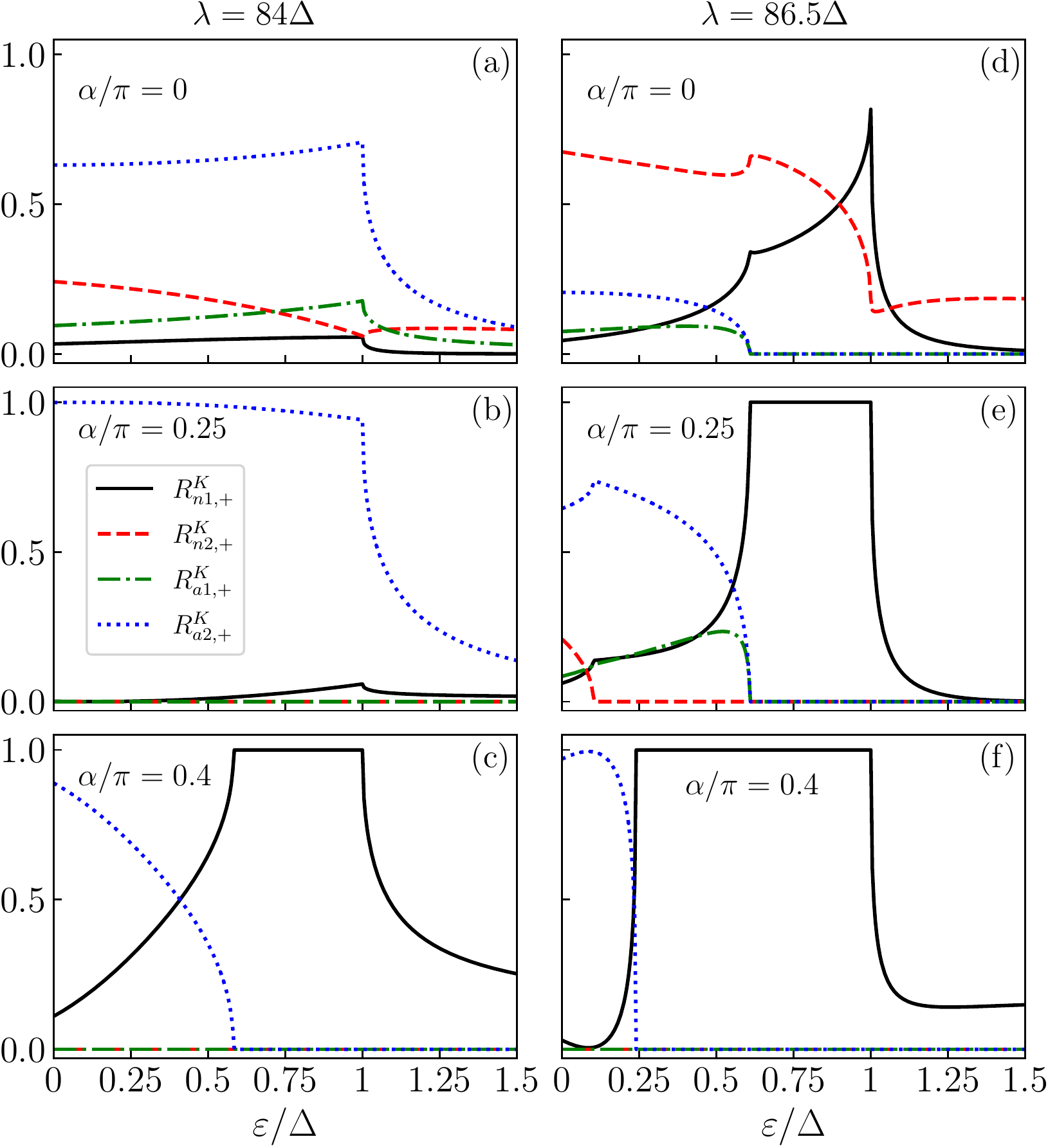}
	\caption{Reflection probabilities (at valley $K$ and $\tau=+$) vs. $\veps$ for $E_F=80\Delta$, $t_\perp=400\Delta$, and $U_0=10\Delta$ with incident angle $\alpha=0$ in (a) and (d), $\alpha=0.25\pi$ in (b) and (e), and $\alpha=0.4\pi$ in (c) and (f). The displacement fields for left and right panels are $\lambda=84\Delta$ and $86.5\Delta$.}
	\label{fig:Reflam80_eps_dep}
\end{figure}
\subsection{$\alpha$-dependent normal and Andreev reflections}
We show the reflection probabilities as a function of $\alpha$ at three $\veps=0.3\Delta$, $0.8\Delta$, and $1.2\Delta$ for $\lambda=84\Delta$ and $86.5\Delta$ in Figs.~\ref{fig:Reflam80_alpha_dep}(a)-(c) and~\ref{fig:Reflam80_alpha_dep}(d)-(f), respectively. Since $\lambda$ is non-zero, these probabilities are slightly asymmetric about $\alpha$. When $\lambda=84\Delta$, the electron and hole excitations span $\veps^{e}_{+,-}\in[-2.55\Delta, \infty]$ and $\veps^{h}_{+,-}\in[-\infty, 2.55\Delta]$; the double normal reflections (SNR and RNR) and double Andreev reflections (SAR and RAR) exist for the intersection of  $-2.55\Delta\lesssim\veps^{e}_{+,-}<4\Delta$ and $-4\Delta<\veps^{h}_{+,-}\lesssim2.55\Delta$. All four probabilities are non-zero only around $\alpha=0$ at $\veps=0.3\Delta$ and $0.8\Delta$. The RAR $R_{a2,+}^K$ is dominating the scattering process. As $\alpha$ is increased from $0$ to $\pi/2$, the $R_{a2,+}^K$ enhances first, reaching to $R_{a2,+}^K\approx 1$, and then starts decreasing and eventually goes to zero, while RNR $R_{n2,+}^K$ and SAR $R_{a1,+}^K$ weaken and slowly vanish, see the results in Figs.~\ref{fig:Reflam80_alpha_dep}(a) and~\ref{fig:Reflam80_alpha_dep}(b). Moreover, the SNR $R_{n1,+}^K$ steadily decreases and almost vanishes before it strikingly rises to the value $1$. In Fig.~\ref{fig:Reflam80_alpha_dep}(c) at $\veps=1.2\Delta$, the magnitude of all reflections is weakened because of the quasiparticles transmission for $\veps>\Delta$. Notice that the sum of all reflection probabilities equals $1$ for $\veps<\Delta$ and becomes less than $1$ for $\veps>\Delta$.

For $\lambda=86.5\Delta$, in Figs.~\ref{fig:Reflam80_alpha_dep}(d)-(f), electron and hole excitations now span $\veps^{e}_{+,-}\in[-0.61\Delta, \infty]$ and $\veps^{h}_{+,-}\in[-\infty, 0.61\Delta]$, and all the four probabilities exist only for the intersection of $-0.61\Delta\lesssim\veps^{e}_{+,-}<6.5\Delta$ and $-6.5\Delta<\veps^{h}_{+,-}\lesssim 0.61\Delta$, see the results in Fig.~\ref{fig:Reflam80_alpha_dep}(d) at $\veps=0.3\Delta$. Interestingly, RAR $R_{a2,+}^K$ is suppressed while RNR $R_{n2,+}^K$ is now dominating the scattering process around $\alpha=0$. For $\veps>0.61\Delta$ in Figs.~\ref{fig:Reflam80_alpha_dep}(e) and~\ref{fig:Reflam80_alpha_dep}(f), the hole excitation $\veps^{h}_{+,-}$ is absent so the double Andreev reflections SAR $R_{a1,+}^K$ and RAR $R_{a2,+}^K$ vanish, but the double normal reflections SNR $R_{n1,+}^K$ and RNR $R_{n2,+}^K$ still exist as the electron excitation $\veps^{e}_{+,-}$ is always present.
\subsection{$\veps$-dependent normal and Andreev reflections}
Here, we discuss these probabilities versus $\veps$, see in Fig.~\ref{fig:Reflam80_eps_dep}, at three incident angles $\alpha=0$, $0.25\pi$, and $0.4\pi$. When $\lambda=84\Delta$, both $\veps^{e}_{+,-}$ and $\veps^{h}_{+,-}$ are present for whole range $0<\veps<1.5\Delta$ therefore all four reflection probabilities are present at $\alpha=0$ [see Fig.~\ref{fig:Reflam80_eps_dep}(a)]. The SNR $R_{n1,+}^K$, SAR $R_{a1,+}^K$, and RAR $R_{a2,+}^K$ are slightly enhanced on increasing $\veps$, and then they are suppressed after $\veps>\Delta$ due to the quasiparticles transmission, whereas the RNR $R_{n2,+}^K$ is opposite in nature. At $\alpha=0.25\pi$ in Fig.~\ref{fig:Reflam80_eps_dep}(b), the RAR starts with $R_{a2,+}^K=1$ as other reflections are zero, and it gradually decreases due to small rise in SNR around $\veps=\Delta$ and then falls rapidly. However, at $\alpha=0.4\pi$, RNR and SAR are always zero while RAR (SNR) decreases (increases) with $\veps$ but RAR vanishes for $\veps<\Delta$ and SNR attains unity, see Fig.~\ref{fig:Reflam80_eps_dep}(c). The critical value of $\veps$ at which RAR vanishes is $\veps_c\approx0.58\Delta$ as the critical angle for retro-reflected hole becomes $\alpha_c\equiv\arcsin{(k^{h+}/k^{e+})}=0.4\pi$ where $k^{h+}$ and $k^{e+}$ are the magnitude of hole and electron momenta.

When we set $\lambda=86.5\Delta$, the hole excitation $\veps^{h}_{+,-}$ is absent for $\veps>0.61\Delta$, and consequently, both Andreev reflections SAR and RAR are zero. However, they are non-zero for $\veps<0.61\Delta$ and are decreasing slowly as $\veps$ is increased, see in Fig.~\ref{fig:Reflam80_eps_dep}(d) for $\alpha=0$. Concerning the double normal reflections, they are non-zero and show non-trivial behaviour. The Fig.~\ref{fig:Reflam80_eps_dep}(e) for $\alpha=0.25\pi$ shows similar results except RNR vanishes before Andreev reflections become zero and SNR attain unity for $0.61\Delta<\veps<\Delta$. For $\alpha=0.4\pi$ in Fig.~\ref{fig:Reflam80_eps_dep}(f), the RAR becomes zero at critical $\veps_c\approx0.24\Delta$, and SNR $R_{n1,+}=1$ for $\veps_c<\veps<\Delta$ as the other two reflections RNR and SAR are always zero. We find that the total probability is equal to $1$ for $\veps<\Delta$, while it is less than $1$ for $\veps>\Delta$.
\bibliography{references_gappedBLG.bib}
\end{document}